\documentclass[11pt,a4paper]{article}
\usepackage{jheppub}
\usepackage{mathtools}
\usepackage{bbold}

\usepackage[english]{babel}

\newmuskip\pFqmuskip

\newcommand*\pFq[6][8]{%
  \begingroup 
  \pFqmuskip=#1mu\relax
  \mathchardef\normalcomma=\mathcode`,
  \mathcode`\,=\string"8000
  \begingroup\lccode`\~=`\,
  \lowercase{\endgroup\let~}\pFqcomma
  {}_{#2}F_{#3}{\left(\genfrac..{0pt}{}{#4}{#5}\Big | #6\right)}
  \endgroup
}
\newcommand{\pFqcomma}{{\normalcomma}\mskip\pFqmuskip}

\newcommand{\pmat}{\begin{pmatrix}}
\newcommand{\fpmat}{\end{pmatrix}}
\newcommand{\eq}{\begin{equation}}
\newcommand{\feq}{\end{equation}}
\newcommand{\cas}{\begin{cases}}
\newcommand{\fcas}{\end{cases}}

\newcommand{\eqarray}{\begin{eqnarray}}
\newcommand{\feqarray}{\end{eqnarray}}

\newcommand{\N}{\mathcal{N}}

\newcommand{\be}{\beta}

\newcommand{\n}{\nu}

\def\n{\nu}									\def\r{\rho}

\def\be{\begin{equation}}
\def\ee{\end{equation}}
\def\bea{\begin{eqnarray}}
\def\eea{\end{eqnarray}}

\title{\centering{Veneziano and Shapiro-Virasoro amplitudes\\ of arbitrarily excited strings}}

\author{Maurizio Firrotta}
\affiliation[a]{Crete Center for Theoretical Physics, Institute for Theoretical and Computational Physics,
\\Department of Physics, Voutes University Campus,
GR-70013, Vasilika Vouton\\ Heraklion, GREECE}

\emailAdd{mfirrotta@physics.uoc.gr}

\abstract{We extend the Veneziano and Shapiro-Virasoro amplitudes to four arbitrarily excited states in bosonic string theory. We use the formalism of coherent string states based on the Di$\,$Vecchia-Del$\,$Giudice-Fubini construction. Within the same formalism, we also analyze the three string scattering finding the covariant version of the three reggeon interaction. Then studying the factorization properties of the extended four string scattering amplitudes we identify the covariant version of the three string interaction. Finally we obtain generalized Kawai-Lewellen-Tye relations connecting the scattering of four open and closed arbitrarily excited states.}

\makeatletter
\gdef\@fpheader{}
\makeatother

\begin{document}
{\flushright CCTP-2024-02\\ \hfill  ITCP-IPP-2024/2}
\maketitle
\section*{Introduction}
String theory's scattering information is very challenging to decode. At the same time its content is also very difficult to falsify, providing in principle a reservoir of non trivial indications. Even without a precise formulation, the scattering properties of string theory emerged in the past, a witness was the dual resonance model (DRM) \cite{Dolen:1967jr}-\cite{Ademollo:1974kz}, the first realization of scattering duality. 
Thereafter, the features of the DRM were recognized and generalized to a model of strings, in particular the three reggeon interaction \cite{Ademollo:1974kz} was identified as the three string interaction \cite{Cremmer:1974jq}, while the four point interactions were classified in terms of their polar structures and high energy behaviors \cite{Veneziano:1968yb}-\cite{Shapiro:1970gy}. Some recent historical recollections can be found in \cite{DiVecchia:2023djc}\cite{Veneziano:2023obw}. The promising characteristics of the hadronic string have been an inspiration for a wide number of important formulations, from the introduction of the fermionic and supersymmetric string \cite{Neveu:1971rx}-\cite{Green:1984sg} to the Heterotic string \cite{Gross:1984dd}-\cite{Candelas:1985en} and many other extensions, including a tadpole free bosonic string \cite{Weinberg:1987ie}. A pedagogical review can be found in \cite{Green:1987sp}\cite{Green:1987mn}.
After the revolutionary work of Friedan-Martinec-Shenker (FMS) \cite{Friedan:1985ge}, in which modern conformal field theory (CFT) techniques were applied to string scattering amplitudes, until nowadays an intensive study of string interactions started pursuing a fundamental theory of particle interactions and quantum gravity. At least to a certain extent, string theory represents an invaluable background where a powerful theoretical technology has been developed and can be applied to challenging physical phenomena, in which a quantum gravity description is needed to reach a deep understanding of the microscopic origin of interaction processes. The black hole (BH) physics provides an example in which a quantum based theory of gravity can shed light on the microscopic description of puzzling phenomena such as chaos, thermal effects and information theory \cite{Gibbons:1977mu}-\cite{Ceplak:2023afb}.
Chasing a microscopic origin of interactions,  the observation of gravitational waves (GW) by ground-based interferometers \cite{LIGOScientific:2016aoc}\cite{LIGOScientific:2016sjg} triggered the quest for a theoretical description of BH collisions \cite{Neill:2013wsa}-\cite{DiVecchia:2023frv} and GW production \cite{Buonanno:2024vkx}-\cite{Barack:2018yly}. Even if the semiclassical approximation seems a useful tool for the description of such interactions, especially in the regime of large distances with respect to the BH size, the characterization of BH scattering including the non point like nature of colliding objects can be tentatively addressed and improved by string inspired frameworks. In particular the string, due to its peculiar degenerate spectrum and its entropic features \cite{Horowitz:1996nw}\cite{Damour:1999aw}, turns out to be a promising candidate for modeling sized effects of heavy massive objects in scattering interactions. Over and above the construction of a string vertex operator containing the full information of the string spectrum \cite{Skliros:2011si}\cite{Aldi:2019osr} and its systematic inclusion in the tree-level string S-matrix \cite{Bianchi:2019ywd} opened a new window on the study of arbitrarily excited string interactions, leading to the observation of erratic behaviors in scattering amplitudes \cite{Gross:2021gsj}\cite{Rosenhaus:2021xhm} until a proposed measure for their chaotic behavior\footnote{More recent studies can be found in \cite{Bianchi:2024fsi}-\cite{Hashimoto:2022bll}  } \cite{Bianchi:2022mhs}-\cite{Bianchi:2023uby}, based on random matrix theory techniques.

The great potential of string theory, that can be employed in a microscopic description of heavy massive objects interactions, and the recently observed features of HES scattering, such as chaos and thermal effects, have motivated us to extend the Veneziano and Shapiro-Virasoro bosonic scattering amplitudes to arbitrarily excited string states.

In the first part of section \ref{sezuno} we briefly review the specific string formalism in order to fix the notation and present the main ingredients that enter in the computation of the scattering amplitudes. In particular we adopt the Di$\,$Vecchia-Del$\,$Giudice-Fubini (DDF) formalism, which combined with the FMS formalism, allows one to build a coherent vertex operator \cite{Skliros:2011si} of string states $i.e.$ a complete object that covers the full BRST cohomology by construction \cite{Brower:1972wj}.

In the second part of section \ref{sezuno} we review the computation of the three reggeon interaction, firstly computed in the context of DRM by \cite{Ademollo:1974kz}, and after some time recognized by \cite{Cremmer:1974jq} to represent the three string interaction. Remarkably we match such computation with the three coherent state computation \cite{Bianchi:2019ywd} and we find a precise agreement with an additional explanation regarding the covariance of the amplitude. In fact, since we find the free scalar parameters of the old fashioned three string vertex determined by scalar product between the DDF light-like reference momenta and the momenta of the three strings, we interpret this feature as a complementary footprint of covariance.

In section \ref{VenHES} we present a systematic way to extend the Veneziano amplitude to arbitrarily excited string states. In particular we find very compact expressions for the classifications of all the contributions coming from the excited nature of the external states, that we parametrize in terms of Jacobi polynomials making manifest the cyclic symmetry of the amplitude. The final expression we find show an exact factorization of the polar structure of the amplitude, represented by the Veneziano factor, from a dressing exponential factor with a suitable combination of Jacobi polynomials, representing the contributions of external states. The computation is also manifestly invariant under the conformal group $SL(2,R)$, in appendix \ref{app1} we report a pedantic derivation of the arbitrarily excited states contributions manipulated respect to the M\"obius transformations. Finally we analyze the factorization properties of the extended Veneziano amplitude finding the correct factorization, which preserves the symmetries, and exposes its structure in terms of the covariant version of three string interactions.

In section \ref{SVsec} we present a systematic extension of the Shapiro-Virasoro amplitude of arbitrarily excited string states. We find a compact expression of the amplitude with a factorized form similar to the one discussed in the open string case. We also obtain a generalized version of the Kawai-Lewellen-Tye (KLT) relations in the context of arbitrarily excited states, which automatically guarantees the correct factorization properties of the closed string amplitude.

Finally, in section \ref{concl} there is a summary of the results and possible future directions.

\section{Covariant DDF states and interactions of three arbitrarily excited strings}\label{sezuno}
In the first part of this introductory section we review the DDF construction of string vertex operators of arbitrarily excited states generated by coherent state techniques. In the second part of this section we review the old fashion three Reggeon vertex, which describes the vertex of three interacting strings, and we outline how the general three point interaction computed in \cite{Bianchi:2019ywd}, based on string coherent state techniques, reproduces the covariant version of such three Reggeon vertex.

\subsection{DDF states and coherent vertex operators}
To start to it, let us introduce the DDF isomorphism of the Heisenberg algebra of creation and annihilation operators
\be\label{isom}
[{ A}_{n}^{i},{ A}_{m}^{j}]=n\,\delta^{ij}\delta_{m{+}n,0}
\ee
where the DDF creation operators in open bosonic string are realized as
\be\label{DDFop}
\lambda_{n}{\cdot}{A}_{-n}=\oint {dz\over 2\pi i}\lambda_{n}{\cdot}i\partial X(z)\, e^{-inq{\cdot}X}(z)
\ee 
with the constraints $\lambda_{n}{\cdot}q{=}0$, $2\alpha'q{\cdot}p_{cm}{=}1$ and $q^{2}{=}0$, where $p_{cm}$ is the center of mass momentum of the string contained in the string operator $X^{j}(z)\sim x_{cm}+p_{cm}\log(z)+\sum_{n\ne 0} \text{osc.}\,z^{n}$. A generic state of the level $N=\sum_{n}ng_{n}$, which corresponds to the mass square $\alpha'M^{2}_{N}=N{-}1$,  is created through the action of DDF creation operators 
\be\label{ms1}
|\{\lambda_{n}\},p_{N}\rangle=\prod_{n}\prod_{\ell=1}^{g_{n}}\lambda^{(\ell)}_{n}{\cdot}{\cal A}_{-n}|\widetilde{p} \rangle
\ee 
on the tachyonic vacuum
\be\label{statecreat}
|\widetilde{p} \rangle\, \Rightarrow\, \lim_{z\rightarrow 0}e^{i\widetilde{p}{\cdot}X}(z)|0\rangle\,, \quad \alpha'\widetilde{p}^{2}=1
\ee
In the pioneer work \cite{Skliros:2011si} it was concretely shown that for every possible state in (\ref{ms1}) one can obtain a covariant version of the corresponding vertex operator by iterating the operator-product-expansion (OPE), such that
 \be
 |\{\lambda_{n}\},p_{N}\rangle=\lim_{z\rightarrow0}{\cal O}\left(\{\lambda_{n}\},\, \partial^{\ell} X\right)e^{i(\widetilde{p}{-}Nq){\cdot}X}(z)|0\rangle
 \ee
 where the operatorial structure $\cal{O}$ is determined by the nature of creation operators, and summarized in the sequence of polarizations $\{\lambda_{n}\}$, while the total momentum of the vertex operator is created by the combination of the tachyonic momentum and DDF reference momenta $q$ specified by the iterative action of creation operators. Since the isomorphism (\ref{isom}) is realized introducing a $q$ dependence in the creation operators, according to (\ref{DDFop}), the total momentum is then given by $p_{N}{=}\widetilde{p}-q\sum_{n}ng_{n}$. An important feature of this formalism is the construction of a coherent vertex operator\footnote{In reference \cite{Skliros:2011si} it was systematically constructed a coherent vertex operator in open and closed bosonic string, while in \cite{Aldi:2019osr} the construction was generalized to superstring theories.} \cite{Skliros:2011si} obtained as a result of the resummed OPE action of a complete exponential of creation operators
\be\label{CoherV}
:e^{\sum_{n}\lambda_{n}{\cdot}{\cal A}_{n}}:\, e^{i\widetilde{p}X}(z)\Big|_{OPE}=\exp{\left(\sum_{n,m}{\zeta_{n}{\cdot}\zeta_{m}\over 2}{\cal S}_{n,m}e^{-i(n{+}m)q{\cdot}X}{+}\sum_{n}\zeta_{n}{\cdot}{\cal P}_{n}e^{-inq{\cdot}X}{+}i\widetilde{p}{\cdot}X\right)}(z)
\ee   
which is still represented in the exponential closed form with only two operatorial structures. The first structure is connected to the term with bilinear polarizations and is given by 
\begin{equation}
S_{m,n} (z)= \sum_{k=1}^n k\, {\cal{Z}}_{n-k}\left( n\frac{\sqrt{2}\ell_{s}q{\cdot}i\partial^s X(z)}{(s{-}1)!}\right) {\cal{Z}}_{m+k}\left( m\frac{\sqrt{2}\ell_{s}q{\cdot}i\partial^s X(z)}{(s{-}1)!}\right) 
\end{equation}
while the second structure is related to the term with a single polarization, which is given by 
\be
  {\cal{P}}^{\mu}_n(z) =\sum_{k=1}^n \frac{ i\partial^k X^{\mu}(z)}{(k{-}1)!} {\cal{Z}}_{n-k}\left( n\frac{\sqrt{2}\ell_{s}q{\cdot}i\partial^s X(z)}{(s{-}1)!}\right)
\end{equation}
both terms are characterized by the cycle index polynomial of the symmetric group, or Schur polynomial\footnote{See \cite{Skliros:2019bqr} for a detailed review the cycle index polynomial properties.  }, defined as
\begin{equation}
{\cal{Z}}_n (a_s)={\cal{Z}}_n (a_1,a_2,..,a_{k})= \oint_{{\cal C}_{0}} \frac{dw}{2\pi iw^{n+1}} \exp\left({-\sum_{s=1}^k \frac{a_s}{s} w^s }\right)
\end{equation}
which is a polynomial of degree $n$ with $k$ elements $a_1,a_2,..,a_{k}$ that in the case of the coherent vertex operator are identified by
\be
a_{s}=n\frac{\sqrt{2}\ell_{s}q{\cdot}i\partial^s X(z)}{(s{-}1)!}
\ee
Given the coherent vertex operator
\be\label{Ocohere}
{\cal V}_{{\cal O}}({\zeta_{n}};z)=\exp{\left(\sum_{n,m}{\zeta_{n}{\cdot}\zeta_{m}\over 2}{\cal S}_{n,m}e^{-i(n{+}m)\sqrt{2}\ell_{s}q{\cdot}X}{+}\sum_{n}\zeta_{n}{\cdot}{\cal P}_{n}e^{-in\sqrt{2}\ell_{s}q{\cdot}X}{+}i\sqrt{2}\ell_{s}\widetilde{p}{\cdot}X\right)}(z)
\ee
with transverse polarizations
\be
\zeta^{\mu}_{n}=\lambda^{\mu}_{n}-\lambda_{n}{\cdot}p\,q^{\mu}
\ee
any vertex operator representing a state of the infinite tower of string states can be selected by the derivative action
\be\label{derproj}
{\cal V}_{\{g_{n}\}}(z)=\prod_{n=1}^{N}\prod_{\ell=1}^{g_{n}}\epsilon_{n}^{(\ell)}{\cdot}{\partial \over \partial \zeta_{n}}{\cal V}_{{\cal O}}({\zeta_{n}};z)
\ee
which enforces the correspondence with partitions of the integer level $N=\sum_{n}ng_{n}$.
The closed string version of such coherent vertex operator is obtained by considering the holomorphic and the anti-holomorphic sectors, with the effects of doubling the operator, and include level matching constraint $N=\overline N$. The resulting operator can be expressed as
\be\label{closedV}
{\cal V}_{{\cal C}}(\zeta_{n},z\,;\overline\zeta_{n},\overline z)={\cal V}_{{\cal O}}({\zeta_{n}};z) \overline{\cal V}_{{\cal O}}({\overline\zeta_{n}};\overline z)
\ee
A pioneer work where tree level scattering amplitudes involving coherent vertex operators were intensively studied is \cite{Bianchi:2019ywd}. The S-matrix elements of coherent states can be used as generating functions of scattering amplitudes. Employing the derivative projection described in (\ref{derproj}), where the polarizations are labeling different states according to (\ref{DDFop}), one can obtain any possible scattering amplitude with desired states.

 In the present work, we will describe the computation of the scattering amplitude with four arbitrarily excited states in open and closed bosonic string theory as a restricted case of the general amplitude involving four coherent states. This is the natural extension of the Veneziano and Shapiro-Virasoro amplitudes where four, in general different, states of the infinite tower of string states interact by tree level scattering.

\subsection{Three arbitrarily excited string interactions }
The old fashion three bosonic string interaction vertex, computed in \cite{Ademollo:1974kz}\cite{Cremmer:1974jq}, was one of the first versions of compact representation of string interactions with the resulting expression written as 
\begin{equation}\label{old3pt}
|V_{\Phi_{1}\Phi_{2}\Phi_{3}}\rangle=\exp \Bigg[ \frac{1}{2}\sum_{\ell,f=1}^{3}\sum_{n,m} (N_{nm}^{(\ell f)})_{Neu}A_{-n}^{(\ell)}{\cdot}A_{-m}^{(f)} + \sum_{\ell=1}^{3}\sum_{n} (N_n^{(\ell)})_{Neu} \sqrt{2}\ell_{s}P^{(\ell+1)}{\cdot}A_{-n}^{(\ell)} \Bigg]|0\rangle
\end{equation}
\begin{figure}[h!]
\centering
\includegraphics[scale=0.5]{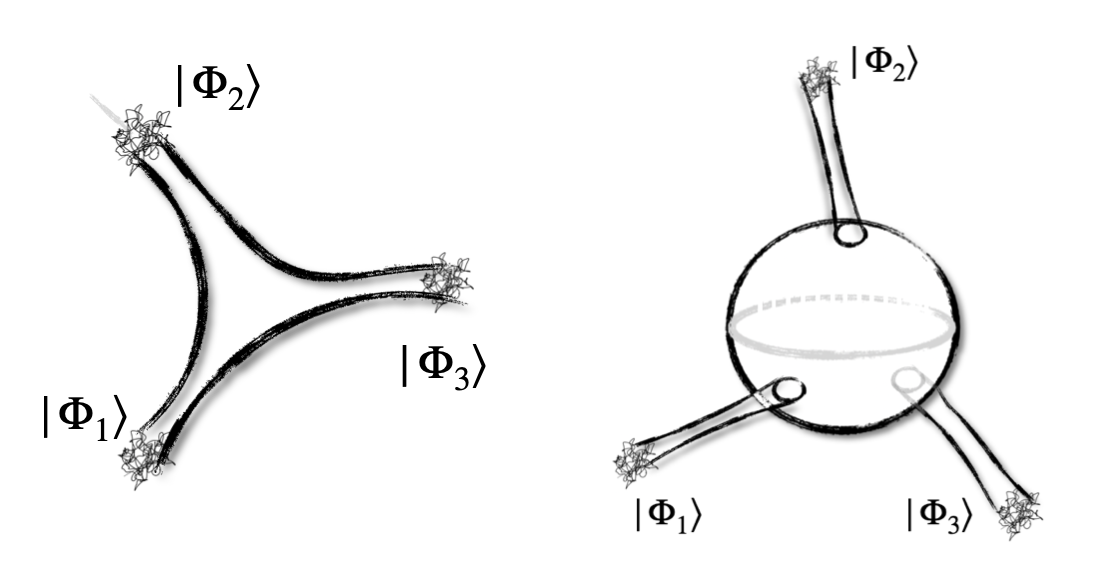}
\caption{Disk and sphere three string interaction vertex.}
\end{figure}
where the polynomial structure is dictated by the Neumann coefficients 
\begin{equation}
(N^{(\ell)}_{n})_{Neu}=-{1\over n\alpha_{\ell{+}1}} \begin{pmatrix}-n{\alpha_{\ell{+}1}\over \alpha_{\ell}}\\n\end{pmatrix}={(-)^{n+1}\over n! \alpha_{\ell}}\left(1+n{\alpha_{\ell+1}\over \alpha_{\ell}}\right)_{n-1}
\end{equation}
\be
(N^{(\ell f)}_{nm})_{Neu} = - \alpha_1 \alpha_2 \alpha_3 \frac{nm}{n\alpha_\ell +m \alpha_f} (N_n^{(\ell)})_{Neu} (N_m^{(f)})_{Neu}
\ee
and the generalized momentum
\begin{equation}
P_{\mu}^{(\ell+1)}=\alpha_{\ell}\,\left(p^{(\ell+1)}_{\mu}-{\alpha_{\ell+1}\over \alpha_{\ell}}p_{\mu}^{(\ell)} \right)
\end{equation}
with an additional consistency constraint involving the three free parameters $\alpha_{1}, \alpha_{2}, \alpha_{3}$  
\be\label{alphapara}
\alpha_1+ \alpha_2+\alpha_3=0
\ee  
this is the three string interaction vertex which is characterized by the Neumann coefficients and the creation operators that can be saturated with any given combination of three strings.

The same interaction can be computed in the string coherent state formalism, where both DDF quantization and vertex operators are combined to produce a covariant result. In fact following \cite{Bianchi:2019ywd} one can use the vertex operators in (\ref{Ocohere}) to compute 
\be
{\cal A}^{3HES}_{gen}(\zeta^{(1)}_{n},\zeta^{(2)}_{n},\zeta^{(3)}_{n})=\Big\langle {\cal V}_{{\cal O}}({\zeta^{(1)}_{n}};z_{1}){\cal V}_{{\cal O}}({\zeta^{(2)}_{n}};z_{2}){\cal V}_{{\cal O}}({\zeta^{(3)}_{n}};z_{3}) \Big\rangle_{D_{2}}
\ee
finding the compact result 
\be\label{3ptGen}
{\cal A}^{3HES}_{gen}(\zeta^{(1)}_{n},\zeta^{(2)}_{n},\zeta^{(3)}_{n})=\exp \Bigg( \sum_{n=1}^\infty \sum_{r=1}^3   \sqrt{2\alpha'} {\cal E}_{n}^{(r)}{\cdot}p_{r+1} +\sum_{n,m=1}^\infty {1\over2}\sum_{rs=1}^{3}  C_{nm}^{r,s} {\cal E}_n^{(r)}{\cdot}{\cal E}_m^{(s)}   \Bigg)
\ee
where the general polarizations are expressed as
\be\label{CcoeffNue}
{\cal E}_{n}^{(\ell)\mu}=\zeta^{(\ell)\mu}_{n}N_{n}^{(\ell)}, \quad \zeta_{n}^{(\ell)\mu}=\lambda_{n}^{(\ell)\mu}-2\ell_{s}^{2}\lambda_{n}^{(\ell)}{\cdot}p_{\ell}\,q_{\ell}^{\mu}    
\ee
with explicit form of the coefficients given by
\begin{equation}\label{CcoeffNue2}
N_n^{(p)}=    \frac{\Gamma ( -2\ell_{s}^{2} n q_p p_{p+1})}{\Gamma (n) 
\Gamma ( -2\ell_{s}^{2}n q_p p_{p+1}-n +1)} = \frac{(-1)^{n{+}1}  ( 1+ n2\ell_{s}^{2}q_p p_{p+1})_{n-1} }{(n-1)!} 
\end{equation}
and
\be
C_{nm}^{rs}=-{n m\over m+n 2\ell_{s}^{2}q_{r}p_{s}} 2\alpha'q_{r}p_{s+1} \, 2\ell_{s}^{2}q_{r}p_{s-1}
\ee
Starting from the first term in (\ref{old3pt})
\begin{equation}
  \sqrt{2 \alpha'}(N_n^{(\ell)})_{Neu} P_{\ell}{\cdot}A^{(\ell)}_{-n}=\sqrt{2\alpha'} ( p^{\mu}_{\ell+1} - \frac{\alpha_{\ell+1}}{\alpha_\ell} p^{\mu}_\ell)   A^{(\ell)}_{\mu,-n}\frac{\Gamma (-n \frac{\alpha_{\ell+1}}{\alpha_\ell})}{n!  \Gamma (-n \frac{\alpha_{\ell+1}}{\alpha_\ell} -n+1)}
\label{AC15} 
\end{equation}
 by matching it with the first term in (\ref{3ptGen}), one can see that the two terms are the same 
 \be\label{momneu}
  \sqrt{2 \alpha'}(N_n^{(\ell)})_{Neu} P_{\ell}{\cdot}A^{(\ell)}_{-n}\Rightarrow  \sqrt{2\alpha'} {\cal E}_{n}^{(r)}{\cdot}p_{r+1}
 \ee
 if one identifies, according to (\ref{ms1}), the creation operators with the DDF polarizations 
 \be
A^{(\ell)}_{\mu,-n}\Rightarrow \lambda_{n,\mu}^{(\ell)}
\ee
but most importantly one can identify the parameters of the old formalism (\ref{alphapara}), in terms of scalar product between the the DDF reference momenta and the momenta of the three strings 
\be
\frac{\alpha_{\ell+1}}{\alpha_\ell} = 2\ell_{s}^{2} q_\ell p_{\ell+1}
\ee
which automatically satisfy the condition (\ref{alphapara}) because it is exactly the momentum conservation. With these identifications one can see from the first term in (\ref{momneu}) how the transverse polarization present in (\ref{CcoeffNue}) is obtained 
\be
 ( p^{\mu}_{\ell+1} - \frac{\alpha_{\ell+1}}{\alpha_\ell} p^{\mu}_\ell)   A^{(\ell)}_{\mu,-n}\Rightarrow \lambda^{(\ell)}_{n}{\cdot}p_{\ell+1} -2\ell_{s}^{2} \lambda_{n}^{(\ell)}{\cdot}p_{\ell}\,q_{\ell+1}{\cdot}p_{\ell+1}=\zeta_{n}^{(\ell)}{\cdot}p_{\ell+1}
\ee
and finally the quadratic term of  (\ref{old3pt}) is precisely reproduced  by 
\be
 (N_{nm}^{\ell v})_{Neu} A_{-n}^{(\ell)}{\cdot}A_{-m}^{v} \Rightarrow C_{nm}^{\ell,v} {\cal E}_n^{(\ell)}{\cdot}{\cal E}_m^{(v)} 
\ee
To summarize, given the identifications
\be
A^{(\ell)}_{\mu,-n}\Rightarrow \lambda_{n,\mu}^{(\ell)}\, ; \quad \frac{\alpha_{\ell+1}}{\alpha_\ell} = 2\ell_{s}^{2} q_\ell p_{\ell+1}
\ee
the scattering amplitude of coherent states (\ref{3ptGen}) and the three string interaction vertex (\ref{old3pt}) matches exactly. This is not very surprising, but quite interestingly the (scalar) parameters of the old fashion computation are in correspondence with scalar products of momenta enforcing the covariance. 

In the next section we will extend the scattering to four arbitrarily excited states and we will find the correct factorization in terms of the generating amplitude of three string interaction. 

\section{Veneziano amplitude of four arbitrary excited strings}\label{VenHES}
In the present section we focus on the computation of the scattering amplitude involving four arbitrarily excited states in open bosonic string theory.
\begin{figure}[h!]
\centering
\includegraphics[scale=0.45]{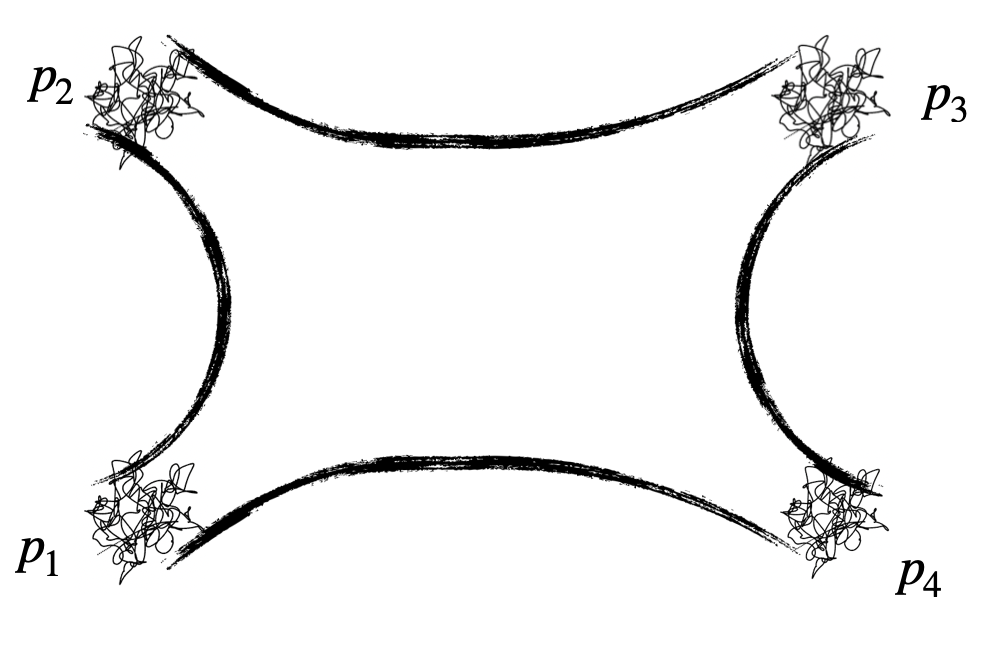}
\caption{Disk amplitude with four HES punctures. }
\end{figure}
\subsection{Four arbitrary excited string interaction}
In this section we present the main steps for the computation of the scattering amplitude of four generic HES states in open bosonic string. Let's start by inserting coherent vertex operators 
\be\label{Ocohere}
{\cal V}_{{\cal O}}({\zeta_{n}};z)=\exp{\left(\sum_{n,m}{\zeta_{n}{\cdot}\zeta_{m}\over 2}{\cal S}_{n,m}e^{-i(n{+}m)\sqrt{2}\ell_{s}q{\cdot}X}{+}\sum_{n}\zeta_{n}{\cdot}{\cal P}_{n}e^{-in\sqrt{2}\ell_{s}q{\cdot}X}{+}i\sqrt{2}\ell_{s}\widetilde{p}{\cdot}X\right)}(z)
\ee
as punctures of the disk in positions $z_{j}$, then the generating scattering amplitude is given by\footnote{We adopted the notation where each vertex operators is weighted by one power of the string length $\ell_{s}$ and from the perturbative expansion in the path integral each puncture takes a power of the open string coupling $g_{o}$. Consequently the disk normalization constant is fixed by unitarity to be $C_{D_{2}}={1\over g_{o}^{2}\ell_{s}^{4}}$.}
\be\label{A4}
{\cal A}_{gen}^{HES}(s,t)=g_{o}^{4}\ell_{s}^{4}C_{D_{2}}\int_{z_{4}}^{z_{2}} \prod_{\ell=1}^{4}dz_{\ell}\,\Big\langle V_{{\cal O}}(p_{1},z_{1})\,V_{{\cal O}}(p_{2},z_{2})\,V_{{\cal O}}(p_{3},z_{3})\,V_{{\cal O}}(p_{4},z_{4}) \Big\rangle_{D_{2}}
\ee
From the correlator one can factorize the Koba-Nielsen contribution due to the contractions 
\be
\langle p_{j}{\cdot}X(z_{j})\,p_{\ell}{\cdot}X(z_{\ell})\rangle=-\,p_{j}{\cdot}p_{\ell}\log(z_{j\ell}) \,, \quad z_{j\ell}=z_{j}{-}z_{\ell}
\ee
finding 
\be
KN(\{z_{j}\})=z_{12}^{2\ell_{s}^{2}p_{1}{\cdot}p_{2}}z_{13}^{2\ell_{s}^{2}p_{1}{\cdot}p_{3}}z_{14}^{2\ell_{s}^{2}p_{1}{\cdot}p_{4}}z_{23}^{2\ell_{s}^{2}p_{2}{\cdot}p_{3}}z_{24}^{2\ell_{s}^{2}p_{2}{\cdot}p_{4}}z_{34}^{2\ell_{s}^{2}p_{3}{\cdot}p_{4}}
\ee
Using the general kinematics
\be
(p_{3}{+}p_{4})^{2}=-s=(p_{1}{+}p_{2})^{2}\,,\,\,\,(p_{2}{+}p_{3})^{2}=-t=(p_{1}{+}p_{4})^{2}\,,\,\,\, (p_{1}{+}p_{3})^{2}=-u=(p_{2}{+}p_{4})^{2}\,,
\ee
where $s+t+u=\sum_{j}M_{N_{j}}^{2}$ with $\ell_{s}^{2}M_{N_{j}}^{2}=N_{j}{-}1$ the mass square of a generic state of the level $N_{j}$ in position $z_{j}$, the Koba-Nielsen contribution is given by
\be\label{KN}
KN(\{z_{j}\})=\left(z_{12}z_{34}\over z_{13}z_{24} \right)^{-{\ell_{s}^{2}s}{-}2}\left(z_{14}z_{23}\over z_{13}z_{24} \right)^{-{\ell_{s}^{2}t}{-}2} (z_{13}z_{24})^{{-}2}\, \prod_{\ell=1}^{4}\left({z_{\ell,\ell{+}1}z_{\ell,\ell{-}1}\over z_{\ell{+}1,\ell{-}1} }\right)^{N_{j}}
\ee
where explicitly the product produces 
\be\label{Nfact}
\prod_{\ell=1}^{4}\left({z_{\ell,\ell{+}1}z_{\ell,\ell{-}1}\over z_{\ell{+}1,\ell{-}1} }\right)^{N_{j}}=\left(z_{12}z_{14}\over z_{24}\right)^{N_{1}}\left(z_{12}z_{23}\over z_{13}\right)^{N_{2}}\left(z_{23}z_{34}\over z_{24}\right)^{N_{3}}\left(z_{14}z_{34}\over z_{13}\right)^{N_{4}}
\ee
with the cyclic identifications $z_{j,4+1}=z_{j,1}$ and $z_{j,1-1}=z_{j,4}$.

The factor (\ref{Nfact}) is related to the exponential nature of the operator insertions (\ref{Ocohere}), and it can be recasted as a dressing factor of the linear and bilinear contributions 
\be
e^{-in_{\ell}\sqrt{2}\ell_{s}q_{\ell}{\cdot}X}\Rightarrow  \left({z_{\ell,\ell+1}z_{\ell,\ell-1}\over z_{\ell+1,\ell-1}} \right)^{n_{\ell}} \,, \quad e^{-i(m_{\ell}+n_{\ell})q_{\ell}{\cdot}X}\Rightarrow  \left({z_{\ell,\ell+1}z_{\ell,\ell-1}\over z_{\ell+1,\ell-1}} \right)^{m_{\ell}+n_{\ell}}
\ee 
this is a rearrangement that makes manifest the mass shell constraint of generic vertex operators, directly inside the scattering amplitude, in fact from the DDF mass shell condition 
\be
\ell_{s}^{2}M_{N_{\ell}}=-\ell_{s}^{2}(\tilde{p}_{\ell}-N_{\ell}q_{\ell})^{2}=-\ell_{s}^{2}\tilde{p}_{\ell}^{2}-2\ell_{s}^{2}N_{\ell}q_{\ell}{\cdot}\tilde{p}_{\ell}=N_{\ell}-1
\ee
fixing the DDF reference vector $q_{\ell}$, one can recover the mass level $N_{\ell}$ trough the inclusion of $n_{\ell}$ DDF vectors $q_{\ell}$ with multiplicity $g_{n_{\ell}}$ so that
\be
N_{\ell}=\sum_{n_{\ell}}g_{n_{\ell}}n_{\ell}
\ee
In order to compute the Wick contractions in (\ref{A4}) a suitable choice, which was pointed out in \cite{Bianchi:2019ywd}, is to take the DDF reference momenta $q_{\ell}$ parallel to each other. With this choice one can start to implement the rule
\be
\langle p_{j}{\cdot}X(z_{j})\,\zeta^{(\ell\ne j)}_{n}{\cdot}\partial X(z_{4}) \rangle={\zeta^{(\ell)}_{n}{\cdot}p_{j}\over z_{j4}}
\ee
that applied to the operatorial structures $\zeta^{(\ell)}_{n}{\cdot}{\cal P}^{(\ell)}_{n}$ and ${\cal S}^{(\ell)}_{n,m}$ yield a pair of contributions for each vertex, one linear in the polarizations
\be
\zeta_{n}^{(\ell)}{\cdot}{\cal P}^{(\ell)}_{n}(z_{\ell})\Rightarrow\, {\cal V}^{(\ell)}_{n}(z_{j,\ell})=-\sum_{k=1}^{n}\left(\sum_{j\ne \ell}{\sqrt{2}\ell_{s}\zeta_{n}^{(\ell)}{\cdot}p_{j}\over z_{j\ell}^{k}}\right){\cal Z}_{n-k}\left(n\sum_{j\ne\ell}{2\ell_{s}^{2}q_{\ell}{\cdot}p_{j}\over z^{f}_{j\ell}} \right)
\ee
and the other one bilinear in the polarizations 
\be
{\cal S}_{n,m}^{(\ell)}(z_{\ell})\Rightarrow\, {\cal W}_{n,m}^{(\ell)}(z_{j\ell})= \sum_{r=1}^{m}r\,{\cal Z}_{n{+}r}\left(n\sum_{j\ne\ell}{2\ell_{s}^{2}q_{\ell}{\cdot}p_{j}\over z^{f}_{j\ell}} \right){\cal Z}_{m{-}r}\left(m\sum_{j\ne\ell}{2\ell_{s}^{2}q_{\ell}{\cdot}p_{j}\over z^{f}_{j\ell}} \right)
\ee 
Finally due to the contractions 
\be
\langle \zeta^{(\ell)}_{n}{\cdot}\partial X(z_{1})\,\zeta^{(f)}_{m}{\cdot}\partial X(z_{4}) \rangle={\zeta^{(\ell)}_{n}{\cdot}\zeta^{(f)}_{m}\over z_{\ell f}^{2}}
\ee
the interaction term between pairs of HES states is obtained as
\be
{\cal I}^{(v,f)}_{n,m}(z_{v f})=\sum_{r=1}^{n}\sum_{\ell=1}^{m} {(-)^{r{+}1}\over z_{v f}^{r{+}\ell}}{\Gamma(r{+}\ell)\over \Gamma(\ell)\Gamma(k)}{\cal Z}_{n{-}r}\left(n\sum_{j\ne v}{2\ell_{s}^{2}q_{v}{\cdot}p_{j}\over z^{h}_{jv}} \right){\cal Z}_{m{-}\ell}\left(m\sum_{j\ne f}{2\ell_{s}^{2}q_{f}{\cdot}p_{j}\over z^{h}_{jf}} \right)
\ee  
Combining all the Wick contractions, the scattering amplitude can be written as
\be\label{intpreVen}
\begin{split}
&{\cal A}_{gen}^{4HES}(s,t)=g_{o}^{2}\int_{z_{4}}^{z_{2}} \prod_{\ell=1}^{4}dz_{\ell}\,\left(z_{12}z_{34}\over z_{13}z_{24} \right)^{-{\ell_{s}^{2}s}{-}2}\left(z_{14}z_{23}\over z_{13}z_{24} \right)^{-{\ell_{s}^{2}t}{-}2} (z_{13}z_{24})^{{-}2}\\
&\exp{\left\{\sum_{\ell=1}^{4}\sum_{n_{\ell}}  \left({z_{\ell,\ell+1}z_{\ell,\ell-1}\over z_{\ell+1,\ell-1}} \right)^{n_{\ell}}\hspace{-2mm} {\cal V}_{n_{\ell}}^{(\ell)}(z_{j\ell}) {+}\sum_{\ell=1}^{4}\sum_{n_{\ell}<m_{\ell}}{\zeta^{(\ell)}_{n_{\ell}}{\cdot}\zeta^{(\ell)}_{m_{\ell}}}  \left({z_{\ell,\ell+1}z_{\ell,\ell-1}\over z_{\ell+1,\ell-1}} \right)^{n_{\ell}+m_{\ell}}\hspace{-4mm} {\cal W}^{(\ell)}_{n_{\ell},m_{\ell}}(z_{j\ell})\right\}}\\
&\exp{\left\{\sum_{v<f=1}^{4} \sum_{n_{v},m_{f}}{\zeta^{(v)}_{n_{v}}{\cdot}\zeta^{(f)}_{m_{f}}}  \left({z_{v,v+1}z_{v,v-1}\over z_{v+1,v-1}} \right)^{n_{v}}\left({z_{f,f+1}z_{f,f-1}\over z_{f+1,f-1}} \right)^{m_{f}} {\cal I}^{(v,f)}_{n_{v},m_{f}}(z_{vf})\right\}}
\end{split}
\ee
and eliminating the redundancy of the $SL(2,\mathit{R})$ invariance by fixing $z_{1},z_{2}$ and $z_{4}$, reflected in the integration measure as
\be
\prod_{j=1}^{4}dz_{j}=z_{12}z_{14}z_{24}\, dz_{3}
\ee
and combining the measure with the factor $(z_{13}z_{24})^{-2}$ in (\ref{KN}), one can see how the result remains finite when $z_{1}{=}\infty,\,z_{2}{=}1,\,z_{3}=z,\,z_{4}{=}0$, since 
\be
{z_{12}z_{14}\over z_{13}^{2}z_{24} }\Big|_{z_{1}=\infty}=1
\ee 
and other terms are 
\be
\left(z_{12}z_{34}\over z_{13}z_{24} \right)^{-{\ell_{s}^{2}s}{-}2}=z^{-{\ell_{s}^{2}s}{-}2}\,, \quad  \left(z_{14}z_{23}\over z_{13}z_{24} \right)^{-{\ell_{s}^{2}t}{-}2}=(1{-}z)^{-{\ell_{s}^{2}t}{-}2}
\ee
Finally following app.\ref{app1}, where a detailed manipulation of all the contraction polynomials is provided, one can write the scattering amplitude in the following compact form 
\be\label{intamp}
\begin{split}
{\cal A}_{gen}^{4HES}(s,t)&=g_{o}^{2}\int_{0}^{1}dz\, z^{-{\ell_{s}^{2}s}{-}2}(1{-}z)^{-{\ell_{s}^{2}t}{-}2}\\
&\exp{\left(\sum_{\ell}\sum_{n_{\ell}}V^{(\ell)}_{n_{\ell}}(z)+{1\over 2}\sum_{\ell}\sum_{n_{\ell},m_{\ell}}W^{(\ell)}_{n_{\ell},m_{\ell}}(z)+{1\over 2}\sum_{v,f}\sum_{n_{v},m_{f}} I_{n_{v},m_{f}}^{(v,f)}(z)\right)}
\end{split}
\ee
where the $s,t$ dependence is also encoded in the scalar products $q_{\ell}{\cdot}p_{j}$ which are fixed by the momentum conservation. The general contributions present in the amplitude can be summarized as follows:
terms linear in the HES polarizations
\be\label{exprV2}
V^{(\ell)}_{n}(z)=\sqrt{2}\ell_{s}\zeta_{n}^{(\ell)}{\cdot}p_{\ell+1}P_{n-1}^{\hspace{0 mm}^{\hspace{0 mm}^{(\alpha_{\ell}^{(n)},\beta_{\ell}^{(n)})}}}\hspace{-9mm}(1{-}2 R_{\ell})+R_{\ell}\sqrt{2}\ell_{s}\zeta_{n}^{(\ell)}{\cdot}p_{\ell+2} P_{n-1}^{\hspace{0 mm}^{\hspace{0 mm}^{(\alpha_{\ell}^{(n)}{+}1,\beta_{\ell}^{(n)})}}}\hspace{-12mm}(1{-}2 R_{\ell})
\ee
terms bilinear in the HES polarizations of the same state  
\be\label{finW}
W^{(\ell)}_{n_{\ell},m_{\ell}}(z)={\zeta^{(\ell)}_{n_{\ell}}{\cdot}\zeta^{(\ell)}_{m_{\ell}}}\sum_{r=1}^{m_{\ell}}r\,P_{n_{\ell}+r}^{\hspace{0 mm}^{\hspace{0 mm}^{(\alpha_{\ell}^{(n_{\ell})}-r,\beta_{\ell}^{(n_{\ell})}-1)}}}\hspace{-16.5mm}(1{-}2 R_{\ell})\,\,P_{m_{\ell}-r}^{\hspace{0 mm}^{\hspace{0 mm}^{(\alpha_{\ell}^{(m_{\ell})}+r,\beta_{\ell}^{(m_{\ell})}-1)}}}\hspace{-17mm}(1{-}2 R_{\ell})
\ee
terms bilinear in HES polarizations of adjacent states
\be\label{Ibv} 
\begin{split}
I^{(\ell,\ell-1)}_{n_{\ell},m_{\ell-1}}(z)=&{\zeta^{(\ell)}_{n_{\ell}}{\cdot}\zeta^{(\ell-1)}_{m_{\ell-1}}}(-)^{m_{\ell-1}{+}1}\sum_{r,s=0}^{n_{\ell},m_{\ell-1}}R_{\ell}^{r+s+1} \\
&P_{n_{\ell}{-}r{-}s{-}1}^{\hspace{0 mm}^{\hspace{0 mm}^{(\alpha_{\ell}^{(n_{\ell})}{+}r{+}s{+}1;\,\beta_{\ell}^{(n_{\ell})}{-}1)}}}\hspace{-16mm}(1{-}2 R_{\ell})\,\,P_{m_{\ell-1}{-}r{-}s{-}1}^{\hspace{0 mm}^{\hspace{0 mm}^{(\beta_{\ell-1}^{(m_{\ell-1})}+r+s+1;\,\alpha_{\ell-1}^{(m_{\ell-1})}-1)}}}\hspace{-20mm}(1{-}2 R_{\ell})
\end{split}
\ee
and finally terms bilinear in HES polarizations of non adjacent states
\be\label{Ibl}
\begin{split}
I^{(\ell,\ell+2)}_{n_{\ell},m_{\ell+2}}(z)=&{\zeta^{(\ell)}_{n_{\ell}}{\cdot}\zeta^{(\ell+2)}_{m_{\ell+2}}}(R_{\ell+1})^{m_{\ell+2}{+}n_{\ell}}\sum_{r,s=0}^{n_{\ell},m_{\ell-1}}\left({R_{\ell}\over R_{\ell+1}}\right)^{r+s+1}\\
&P_{n_{\ell}{-}r{-}s{-}1}^{\hspace{0 mm}^{\hspace{0 mm}^{(\alpha_{\ell}^{(n_{\ell})}{+}r{+}s{+}1;\,\sigma_{\ell}^{(n_{\ell})}{-}1)}}}\hspace{-16mm}(1{+}2 {R_{\ell}/ R_{\ell+1}})\,\,P_{m_{\ell+2}{-}r{-}s{-}1}^{\hspace{0 mm}^{\hspace{0 mm}^{(\alpha_{\ell+2}^{(m_{\ell+2})}+r+s+1;\,\sigma_{\ell+2}^{(m_{\ell+2})}-1)}}}\hspace{-20mm}(1{+}2 R_{\ell}/R_{\ell+1})
\end{split}
\ee
with coefficients
\be
R_{1}={z_{21}z_{34}\over z_{24}z_{31}}=z\,,\quad R_{2}={z_{32}z_{41}\over z_{31}z_{42}}=1{-}z\,,\quad R_{3}={z_{43}z_{12}\over z_{42}z_{13}}=z\,,\quad R_{4}={z_{14}z_{23}\over z_{13}z_{24}}=1{-}z
\ee
and
\be\label{coeffJac}
\alpha_{\ell}^{(n)}=-n-2\ell_{s}^{2}nq_{\ell}{\cdot}p_{\ell+1}\,,\quad \beta_{\ell}^{(n)}=-n-2\ell_{s}^{2}nq_{\ell}{\cdot}p_{\ell-1}\,,\quad  \sigma_{\ell}^{(n)}=-n-2\ell_{s}^{2}n q_{\ell}{\cdot}p_{\ell+2}
 \ee
where every contribution is expressed in terms of Jacobi polynomials 
\be
P_{N}^{(\alpha,\beta)}(x)=\sum_{r=0}^{N}\begin{pmatrix}N+\alpha\\N-r\end{pmatrix}\begin{pmatrix}N+\beta\\r\end{pmatrix}\left({x-1\over 2}\right)^{r}\left({x+1\over 2}\right)^{N-r}
\ee
 
Using the formula
\be\label{formVen}
e^{\sum_{n}a_{n}z^{n}}e^{\sum_{n}b_{n}(1{-}z)^{n}}=\exp\left(\sum_{n}a_{n}\partial^{n}_{\beta_{s}}\right)\exp\left(\sum_{n}b_{n}\partial^{n}_{\beta_{t}}\right)e^{\beta_{s}z+\beta_{t}\,(1{-}z)}\Big|_{\beta_{s,t}=0}
\ee
all the polynomial dependence on $z$ and $1{-}z$ of the functions $V_{n}^{(\ell)}$, $W_{n,m}^{(\ell,\ell)}$ and $I_{n,m}^{(v,f)}$  can be identified with the derivatives respect to $\beta_{s}$ and $\beta_{t}$ as
\be\label{idVen}
z^{k}\rightarrow \partial^{k}_{\beta_{s}}\,,\quad (1{-}z)^{k}\rightarrow \partial^{k}_{\beta_{t}}
\ee
and the expression (\ref{intamp}) can be integrated, finding a more compact representation provided by
\be
{\cal A}_{gen}^{4HES}(s,t)={\cal A}_{Ven}(s,t)\,e^{{\cal K}\left(\{\zeta^{(\ell)}_{n}\};\partial_{\beta_{s}},\partial_{\beta_{t}}\right)}\Phi_{\beta_{s},\beta_{t}}\left(s,t\right)\Big|_{\beta_{s,t}{=}0}\ee
This result makes manifest the factorization of a term which contains all the poles structure, a Veneziano-like term of the form 
\be
{\cal A}_{Ven}(s,t)=g_{o}^{2}{\Gamma({-}\ell_{s}^{2}s{-}1)\Gamma({-}\ell_{s}^{2}t{-}1)\over \Gamma({-}\ell_{s}^{2}s{-}\ell_{s}^{2}t{-}2)}
\ee
from the exponential dressing factor with argument 
\be\label{Kfunc}
\begin{split}
{\cal K}\left(\{\zeta^{(\ell)}_{n}\}; \partial_{\beta_{s}},\partial_{\beta_{t}}\right)=&\sum_{\ell=1}^{4}\left(\sum_{n}V^{(\ell)}_{n}(\partial_{\beta_{s,t}}){+}\sum_{n,m} W_{n,m}^{(\ell)}(\partial_{\beta_{s,t}})\right)+\sum_{v<f=1}^{4}\sum_{n,m} I_{n,m}^{(v,f)}(\partial_{\beta_{s,t}})
\end{split}
\ee
which encodes the non trivial structure of all the possible external HES states. Any precise structure of the four HES will emerge from the polynomial action on the pole-free function
\be\label{ffunc}
\Phi_{\beta_{s},\beta_{t}}(s,t)=\sum_{r=0}^{\infty}\sum_{v=0}^{\infty}{\beta_{s}^{r}\over r!}{\beta_{t}^{v}\over v!}{(-\ell_{s}^{2}s{-}1)_{r}(-\ell_{s}^{2}t{-}1)_{v}\over (-\ell_{s}^{2}s{-}\ell_{s}^{2}t{-}2)_{r+v}}
\ee
that will systematically generate a combination of contributions according to    
\be
\partial^{a}_{\beta_{s}}\partial^{b}_{\beta_{s}}\Phi_{\beta_{s},\beta_{t}}(s,t)\Big|_{\beta_{s,t}{=}0}={(-\ell_{s}^{2}s{-}1)_{a}(-\ell_{s}^{2}t{-}1)_{b}\over (-\ell_{s}^{2}s{-}\ell_{s}^{2}t{-}2)_{a+b}}
\ee  
Following the analogy with (\ref{derproj}), one can generate any desired amplitude with specific external HES states by taking the derivative projection 
\be\label{Mfor}
{\cal A}_{N_{1},N_{2},N_{3},N_{4}}^{4HES}(s,t)=\prod_{\ell=1}^{4}\prod_{n_{\ell}=1}^{N_{\ell}}{1\over \sqrt{n_{\ell}^{g_{n_{\ell}}}g_{n_{\ell}}!}}\prod_{a=1}^{g_{n_{\ell}}}\epsilon_{a,n}^{(\ell)}{\cdot}{d\over d\zeta_{n_{\ell}}^{(\ell)}}{\cal A}_{gen}^{HES}(s,t)\Bigg|_{\{\zeta_{n_{\ell}\}}=0}
\ee
which produce the scattering amplitude involving four arbitrary HES with excitation level $N_{\ell}$ and maximal helicity ${\cal J}_{\ell}$ according to the relations
\be
N_{\ell}=\sum_{n_{\ell}=1}^{N_{\ell}}n_{\ell}g_{n_{\ell}}\,, \quad {\cal J}_{\ell}=\sum_{n_{\ell}=1}^{N_{\ell}}g_{n_{\ell}}
\ee
where any precise HES microstate of the level $N_{\ell}$ is identified by the set of coefficients $\{g_{n_{\ell}}\}$. The full amplitude in the abelian case is given by three disk diagrams
\be
{\cal A}_{N_{1},N_{2},N_{3},N_{4}}^{4HES}(s,t,u)={\cal A}_{N_{1},N_{2},N_{3},N_{4}}^{4HES}(s,t)+{\cal A}_{N_{1},N_{2},N_{3},N_{4}}^{4HES}(t,u)+{\cal A}_{N_{1},N_{2},N_{3},N_{4}}^{4HES}(s,u)
\ee
and all the other diagrammatic contributions can be obtained by the symmetry of the disk.

In the next part we will show how the present embedding of the amplitude is manifestly factorizable thanks to the properties of the Jacobi polynomials.
\subsection{Factorization properties}
As a warm-up example one can start by considering the integral representation of $(s,t)$ diagram of the Veneziano amplitude in the bosonic string case \be\label{VenInt}
{\cal A}(s,t)=g_{o}^{2}\int_{0}^{1}dz\, z^{-\ell_{s}^{2}s-2}(1{-}z)^{-\ell_{s}^{2}t-2},
\ee
the s-channel poles factorization can be performed by rewriting the amplitude as
\be
{\cal A}(s,t)=g_{o}^{2}\int_{0}^{1}dz\, z^{-\ell_{s}^{2}s-2} e^{-(\ell_{s}^{2}t{+}2)\log(1{-}z)}
\ee
and considering a formal expansion of the integrand around $z\sim 0$, which leads to 
\be
{\cal A}(s,t)=g_{o}^{2}\int_{0}^{1}dz\, z^{-\ell_{s}^{2}s-2} \exp\left({(\ell_{s}^{2}t{+}2)\sum_{r=1}^{\infty}{z^{r}\over r}}\right)
\ee
one can use the cycle index polynomial to write the amplitude as
\be
{\cal A}(s,t)=g_{o}^{2}\int_{0}^{1}{dz\over z}\, z^{-\ell_{s}^{2}s-1}\sum_{N=0}^{\infty}{\cal Z}_{N}\left(\ell_{s}^{2}t{+}2\right)z^{N}
\ee
and finally computing the integral one gets 
\be
{\cal A}(s,t)=g_{o}^{2}\sum_{N=0}^{\infty} {{\cal Z}_{N}(\ell_{s}^{2}t{+}2)\over {-}\ell_{s}^{2}s{-}1{+}N }
\ee
which give a pole every time $\ell_{s}^{2}s{-}1$ is equal to the integer $N$, with residue
\be
{\cal Z}_{N}(\ell_{s}^{2}t{+}2)={(\ell_{s}^{2}t{+}2)_{N}\over N!}
\ee
This is only an alternative way of writing the pole expansion of the Veneziano amplitude, where we have represented the residues in terms of cycle index polynomials
\be
{\cal A}(s,t)= g_{o}^{2}\sum_{N=0}^{\infty} {{\cal R}_{N}(t)\over s-M_{N}^{2}}
\ee
where exactly
\be
{\cal R}_{N}(t)=-{1\over \ell_{s}}{\cal Z}_{N}(\ell_{s}t{+}2)=-{1\over \ell_{s}}{(\ell_{s}^{2}t{+}2)_{N}\over N!}\,;\quad M_{N}^{2}={(N{-}1)\over \ell_{s}^{2}}
\ee
This observation can be used to show how the combinatorics of the cycle index polynomial put in correspondence the residues with the corresponding states, including the degeneracy. In fact considering for example the cycle index polynomial of two elements $a_{1},a_{2}$, which is given by
\be
{\cal Z}_{n}(a_{1},a_{2})=\oint {dw\over2\pi i w^{1+n}}\exp\left(a_{1}w + {a_{2}\over 2}w^{2}\right)={1\over n!}\partial_{w}^{n}\exp\left(a_{1}w + {a_{2}\over 2}w^{2}\right)\Big|_{w=0}
\ee 
and taking $n=2$ one gets the following combination
\be
{\cal Z}_{2}(a_{1},a_{2})={a_{1}^{2}\over 2!}+{a_{2}\over 2}
\ee
Considering three elements, one gets 
\be
{\cal Z}_{n}(a_{1},a_{2},a_{3})={1\over n!}\partial_{w}^{n}\exp\left(a_{1}w + {a_{2}\over 2}w^{2}+{a_{3}\over 3}w^{3}\right)\Big|_{w=0}
\ee 
that for the specific value of $n=3$ gives
\be
{\cal Z}_{3}(a_{1},a_{2},a_{3})={a_{1}^{3}\over 3!} + {a_{1}a_{2}\over 2 }+{a_{3}\over 3}
\ee
Iterating this procedure one can see that all the combinations obtained from the cycle index polynomials, with degree equal to the number of elements, are in one to one correspondence with the partitions of integers and reproduce the exact systematics of the creation operators action including the normalization. In particular taking the identification $a_{n}\equiv A_{-n}$, $i.e$ each element is a creation operator, it is very easy to see that for the excitation level $N{=}2$ one has two possible actions of creation operators
\be
\left({1\over 2}{A_{-1}A_{-1}} + {A_{-2}\over 2}\right) |0\rangle={\cal Z}_{2}(A_{-1},A_{-2})|0\rangle
\ee
while at $N{=}3$ one has 
\be
\left({A_{-1}^{3}\over 3!} + {A_{-1}A_{-2}\over 2 }+{A_{-3}\over 3}\right)|0\rangle={\cal Z}_{3}(A_{-1},A_{-2},A_{-3})|0\rangle
\ee
and in general 
\be
\sum_{g_{n}:\sum_{n=1}^{N}ng_{n}=N}\prod_{n=1}^{N}{A_{-n}^{g_{n}}\over \sqrt{n^{g_{n}}g_{n}!}}\,|0\rangle={\cal Z}_{N}\Big(\{A_{-j}\}_{j=1}^{N}\Big)|0\rangle
\ee
Matching this procedure with the residues, expressed in terms of cycle index polynomials, one can write
\be
{\cal R}_{N}(t)=-{1\over \ell_{s}^{2}} {\cal Z}_{N}(\ell_{s}^{2}t{+}2)=-{1\over \ell_{s}^{2}}{(\ell_{s}^{2}t{+}2)_{N}\over N!}=-{1\over \ell_{s}^{2}}{1\over N!}\partial_{w}^{N}\exp\left((\ell_{s}t{+}2)\sum_{\ell=0}^{\infty}{w^{\ell}\over \ell} \right)\Big|_{w=0}
\ee
and introducing the set of parameters $a_{n}$ in the residues as 
\be
\widehat{\cal R}_{N}(t;\{a_{n}\})=-{1\over \ell_{s}^{2}} {\cal Z}_{N}(\ell_{s}^{2}t{+}2;\{a_{n}\}_{1}^{N})=-{1\over \ell_{s}^{2}}{1\over N!}\partial_{w}^{N}\exp\left((\ell_{s}t{+}2)\sum_{n=0}^{N}{a_{n}w^{n}\over n} \right)\Big|_{w=0}
\ee
one can read the fragmentation of the residue in terms of the states at each excitation level $N$ by identifying the set $\{a_{n}\}$ with avatar coefficients of creation operators. In this setup the residue fragmentation in terms of states can be written as 
\be
{\cal R}_{N=2}(t)={(\ell_{s}^{2}t{+}2)^{2}\over 2}+{(\ell_{s}^{2}t{+}2)\over 2}\Rightarrow \widehat{\cal R}_{N=2}(t;A_{-1},A_{-2})={(\ell_{s}^{2}t{+}2)^{2}\over 2}A_{-1}A_{-1}+{(\ell_{s}^{2}t{+}2)\over 2}A_{-2}
\ee
and the same can be repeated for each excitation level. A natural filter for the identification of the states in the pole expansion of the Veneziano amplitude is provided by the cycle index polynomial, with the resulting modification 
\be
{\cal A}_{Ven}^{filt}(s,t)=\sum_{N=0}^{\infty}{\widehat{\cal R}_{N}(t;\{A_{-n}\}_{1}^{N})\over s-M_{N}^{2} } \xRightarrow[]{\{A_{-n}\}=1} \sum_{N=0}^{\infty}{{\cal R}_{N}(t)\over s-M_{N}^{2} }={\cal A}_{Ven}(s,t)
\ee
The same analysis can be done for the $t-$channel poles studying the integrand of (\ref{VenInt}) around $z\sim1$.

The non trivial combinatorics of the cycle index polynomial reproduce the exact identification of the states of each level not only in the construction of the vertex operators of arbitrarily excited states but also in the residue fragmentation.

The scattering of four arbitrarily excited states is the result of the polynomial properties of the cycle index, that is representative of the features of the string spectrum, and the properties of the conformal invariance. Their merging produces a suitable polynomial representation of all the scattering contributions given in term of Jacobi polynomials. In order to study the factorization properties of the 4-HES scattering one can start from expression (\ref{intamp})
\be
\begin{split}
{\cal A}_{gen}^{4HES}(s,t)&=g_{o}^{2}\int_{0}^{1}dz\, z^{-{\ell_{s}^{2}s}{-}2}(1{-}z)^{-{\ell_{s}^{2}t}{-}2}\\
&\exp{\left(\sum_{\ell}\sum_{n_{\ell}}V^{(\ell)}_{n_{\ell}}(z)+{1\over 2}\sum_{\ell}\sum_{n_{\ell},m_{\ell}}W^{(\ell)}_{n_{\ell},m_{\ell}}(z)+{1\over 2}\sum_{v,f}\sum_{n_{v},m_{f}} I_{n_{v},m_{f}}^{(v,f)}(z)\right)}
\end{split}
\ee
where the general form of the terms in the exponential is summarized from (\ref{exprV2}) to (\ref{Ibl}) while the explicit expressions are classified in  app.\ref{ClassTerms}, and looking at the $s$-channel factorization one can write 
\be
\begin{split}
{\cal A}_{gen}^{4HES}(s,t)&=g_{o}^{2}\int_{0}^{1}{dz\over z}\, z^{-\ell_{s}^{2}s-1}\sum_{N=0}^{\infty}{\cal Z}_{N}\left(\ell_{s}^{2}t{+}2\right)z^{N}\\
&\exp{\left(\sum_{\ell}\sum_{n_{\ell}}V^{(\ell)}_{n_{\ell}}(z)+{1\over 2}\sum_{\ell}\sum_{n_{\ell},m_{\ell}}W^{(\ell)}_{n_{\ell},m_{\ell}}(z)+{1\over 2}\sum_{v,f}\sum_{n_{v},m_{f}} I_{n_{v},m_{f}}^{(v,f)}(z)\right)}
\end{split}
\ee
which produces the $s$-channel factorization at $\ell_{s}^{2}s=N-1$ of the 4-HES scattering amplitude
\be
\begin{split}
&{\cal A}_{gen}^{4HES}(s,t)=\\
&g_{o}^{2}\sum_{N=0}^{\infty} {{\cal Z}_{N}(\ell_{s}^{2}t{+}2)\over {-}\ell_{s}^{2}s{-}1{+}N }\exp{\left(\sum_{\ell}\sum_{n_{\ell}}V^{(\ell)}_{n_{\ell}}(0)+{1\over 2}\sum_{\ell}\sum_{n_{\ell},m_{\ell}}W^{(\ell)}_{n_{\ell},m_{\ell}}(0)+{1\over 2}\sum_{v,f}\sum_{n_{v},m_{f}} I_{n_{v},m_{f}}^{(v,f)}(0)\right)}
\end{split}
\ee
Now starting from the polynomial $V^{(\ell)}_{\ell}(z)$ one can see that the term in position one can be written as
\be
\begin{split}
V^{(1)}_{n}(z{=}0)&=\sqrt{2}\ell_{s}\zeta_{n}^{(1)}{\cdot}p_{2}P_{n-1}^{\hspace{0 mm}^{\hspace{0 mm}^{(\alpha_{1}^{(n)},\beta_{1}^{(n)})}}}\hspace{-9mm}(1{-}2 z)+z\sqrt{2}\ell_{s}\zeta_{n}^{(1)}{\cdot}p_{3} P_{n-1}^{\hspace{0 mm}^{\hspace{0 mm}^{(\alpha_{1}^{(n)}{+}1,\beta_{1}^{(n)})}}}\hspace{-12mm}(1{-}2z)\Big|_{z=0}\\
&=\sqrt{2}\ell_{s}\zeta_{n}^{(1)}{\cdot}p_{2}P_{n-1}^{\hspace{0 mm}^{\hspace{0 mm}^{(\alpha_{1}^{(n)},\beta_{1}^{(n)})}}}\hspace{-9mm}(1)=\sqrt{2}\ell_{s}\zeta_{n}^{(1)}{\cdot}p_{2} \begin{pmatrix}n-1+\alpha_{1}^{(n)} \\ n-1\end{pmatrix}
\end{split}
\ee
where in the last equality we used the property 
\be
P_{n}^{(a,b)}(1)= \begin{pmatrix}n+a \\ n\end{pmatrix}
\ee
Finally using the expression $\alpha_{1}^{(n)}={-}n{-}2\ell_{s}^{2}nq_{1}{\cdot}p_{2}$ one can write
\be
V^{(1)}_{n}(0)=\sqrt{2}\ell_{s}\zeta_{n}^{(1)}{\cdot}p_{2} \begin{pmatrix} -2\ell_{s}^{2}nq_{1}{\cdot}p_{2}-1 \\ n-1\end{pmatrix}=(-)^{n+1}\sqrt{2}\ell_{s}\zeta_{n}^{(1)}{\cdot}p_{2}{(1{+}2\ell_{s}^{2}nq_{1}{\cdot}p_{2})_{n-1}\over \Gamma(n)}
\ee
where the binomial term is exactly the combination of the Neumann coefficient, with the momentum contribution of position one, that characterize the three point amplitude (\ref{CcoeffNue}) and (\ref{CcoeffNue2}).

From position two, using (\ref{exprze2}), one gets
\be
V^{(2)}_{n}(0)=\sqrt{2}\ell_{s}\zeta_{n}^{(2)}{\cdot}p_{3}P_{n-1}^{\hspace{0 mm}^{\hspace{0 mm}^{(\alpha_{2}^{(n)},\beta_{2}^{(n)})}}}\hspace{-9mm}({-}1)+\sqrt{2}\ell_{s}\zeta_{n}^{(2)}{\cdot}p_{4} P_{n-1}^{\hspace{0 mm}^{\hspace{0 mm}^{(\alpha_{2}^{(n)}{+}1,\beta_{2}^{(n)})}}}\hspace{-12mm}({-}1)
\ee
and using the property
\be
P_{n}^{(a,b)}(-1)=(-)^{n}P_{n}^{(b,a)}(1)= (-)^{n}\begin{pmatrix}n+b \\ n\end{pmatrix}
\ee
with the explicit value of $\beta_{2}^{(n)}{=}{-}n{-}2\ell_{s}^{2}nq_{2}{\cdot}p_{1}$ one finds
\be
V^{(2)}_{n}(0)=(-)^{n}\sqrt{2}\ell_{s}\zeta_{n}^{(2)}{\cdot}p_{1}\begin{pmatrix}-1-2\ell_{s}^{2}nq_{2}{\cdot}p_{1} \\ n-1\end{pmatrix}=-\sqrt{2}\ell_{s}\zeta_{n}^{(2)}{\cdot}p_{1}{(1{+}2\ell_{s}^{2}nq_{2}{\cdot}p_{1})_{n-1}\over \Gamma(n)}
\ee
that reproduces exactly the Neumann coefficient, with the momentum contribution in position two, of the three point amplitude shifted by momentum conservation to the contraction with $p_{1}$ instead of $p_{3}$. In order to see the cyclic structure one can use the momentum conservation of the factorized amplitude $p_{1}+p_{2}+P_{N}=0$ finding 
\be
V^{(2)}_{n}(0)=(-)^{n+1}\sqrt{2}\ell_{s}\zeta_{n}^{(2)}{\cdot}P_{N}{(1{+}2\ell_{s}^{2}nq_{2}{\cdot}P_{N})_{n-1}\over \Gamma(n)}
\ee
and iterating this procedure for the remaining terms one gets
\begin{equation}
V^{(\ell)}_{n}(0)=N_n^{(\ell)}= \frac{(-1)^{n{+}1}  ( 1+ n2\alpha'q_\ell p_{\ell+1})_{n-1} }{(n-1)!} 
\end{equation}
The same kind of considerations can be used to study the bilinear terms leading to the resulting expressions 
\be
W^{(\ell)}_{n,m}(0)=\zeta^{(\ell)}_{n}{\cdot}\zeta^{(\ell)}_{m}C_{n,m}^{\ell,\ell}N_n^{(\ell)}N_m^{(\ell)}\,,\quad {I}^{(\ell,\ell-1)}_{n_{\ell},m_{\ell-1}}(0)={\zeta^{(\ell)}_{n_{\ell}}{\cdot}\zeta^{(\ell-1)}_{m_{\ell-1}}}C_{n_{\ell},m_{\ell-1}}^{\ell,\ell-1}N_{n_{\ell}}^{(\ell)}N_{m_{\ell-1}}^{(\ell-1)}
\ee
which are the covariant version of the Neumann coefficients described in (\ref{CcoeffNue}) and (\ref{CcoeffNue2}). The last two bilinear terms represented by the mixed contractions of non adjacent states (\ref{I13}) and (\ref{I24}) are identically zero
\be
{I}^{(\ell,\ell+2)}_{n_{\ell},m_{\ell+2}}(0)=0
\ee
The result of the factorization can be summarized as 
\be
{\cal A}_{gen}^{4HES}(s,t)\Big|_{P_{N}^{2}\sim N-1}=g_{o}^{2}{\cal A}_{gen}^{2HES + N}(p_{1},p_{2},P_{N}) {{\cal Z}_{N}(\ell_{s}^{2}t{+}2)\over {-}\ell_{s}^{2}P_{N}^{2}{-}1{+}N } {\cal A}_{gen}^{2HES + N}(p_{3},p_{4},-P_{N})
\ee
\begin{figure}[h!]
\centering
{\vspace{-3mm}\includegraphics[scale=0.35]{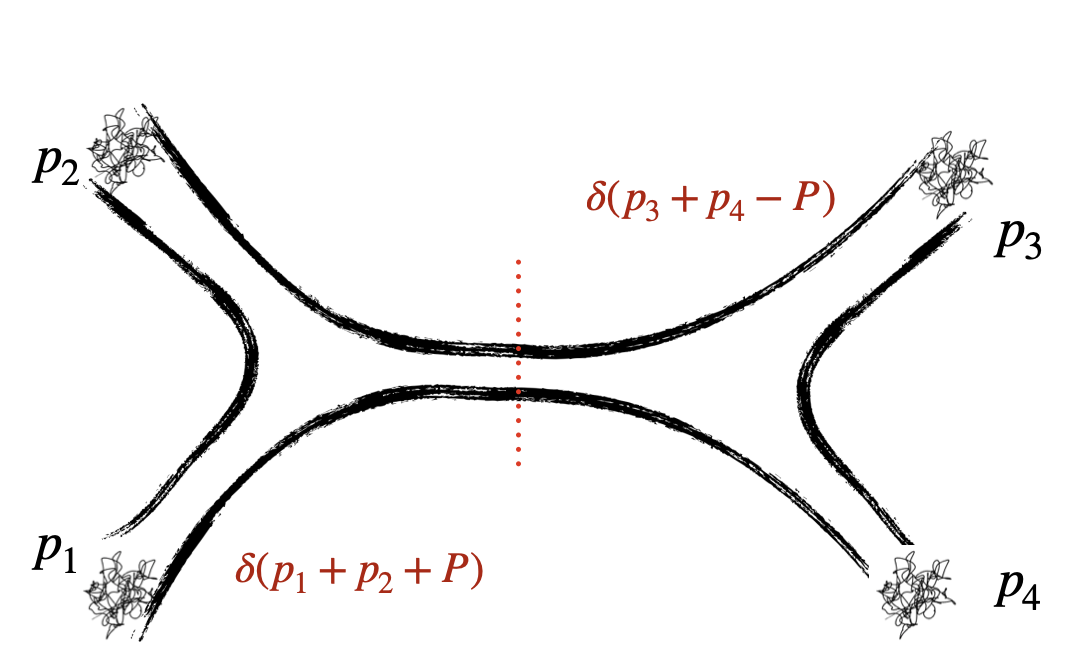}}\quad \quad\quad  \includegraphics[scale=0.3]{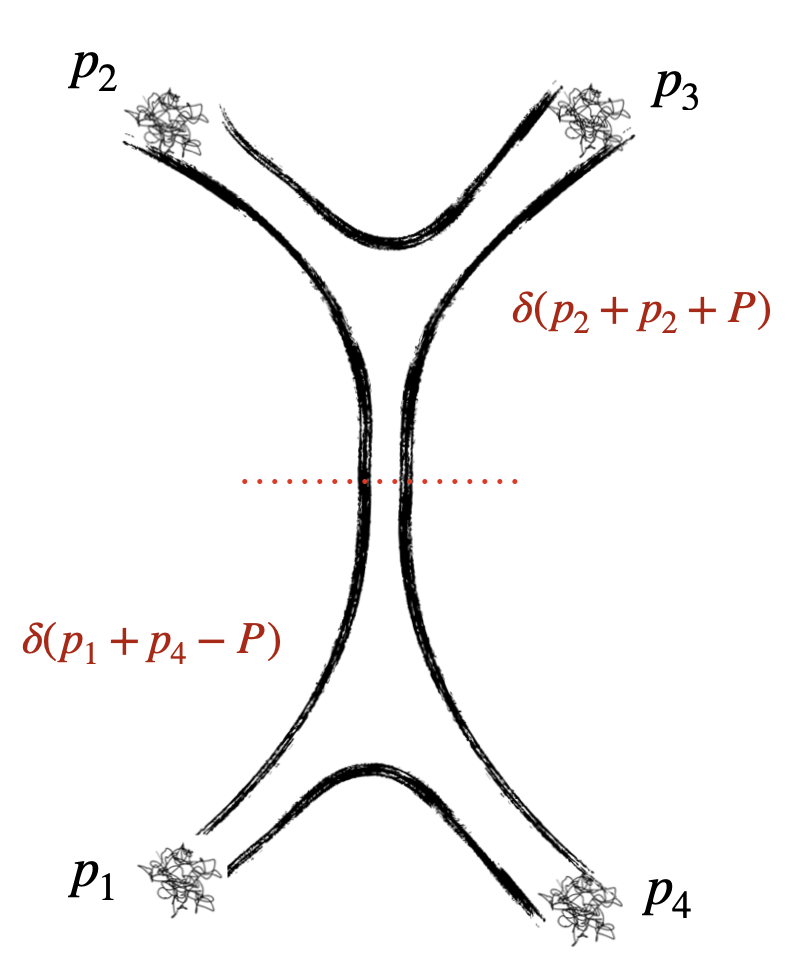}
\caption{Picture of the $s$-channel and $t$-channel factorization.}
\end{figure}
where the three point amplitudes are the generating amplitudes in (\ref{3ptGen}) specialized to the case of two arbitrary states and one internal states given by
\be
\begin{split}
{\cal A}_{gen}^{2HES + N}(p_{1},p_{2},P_{N})=&\exp \Bigg( \sum_{n=1}^\infty  \sqrt{2\ell_{s}} {\cal E}_{n}^{(1)}{\cdot}p_{2}+\sqrt{2\ell_{s}} {\cal E}_{n}^{(2)}{\cdot}P_{N} +\sum_{n,m=1}^\infty  C_{nm}^{1,2} {\cal E}_n^{(1)}{\cdot}{\cal E}_m^{(2)}   \Bigg)\\
&\exp\left( \sum_{n,m=1}^\infty  C_{nm}^{1,1} {\cal E}_n^{(1)}{\cdot}{\cal E}_m^{(1)}+ \sum_{n,m=1}^\infty  C_{nm}^{2,2} {\cal E}_n^{(2)}{\cdot}{\cal E}_m^{(2)} \right)
\end{split}
\ee
and
\be
\begin{split}
{\cal A}_{gen}^{2HES + N}(p_{3},p_{4},-P_{N})=&\exp \Bigg( \sum_{n=1}^\infty  \sqrt{2\ell_{s}} {\cal E}_{n}^{(3)}{\cdot}p_{4}-\sqrt{2\ell_{s}} {\cal E}_{n}^{(4)}{\cdot}P_{N} +\sum_{n,m=1}^\infty  C_{nm}^{3,4} {\cal E}_n^{(3)}{\cdot}{\cal E}_m^{(4)}   \Bigg)\\
&\exp\left( \sum_{n,m=1}^\infty  C_{nm}^{3,3} {\cal E}_n^{(3)}{\cdot}{\cal E}_m^{(3)}+ \sum_{n,m=1}^\infty  C_{nm}^{4,4} {\cal E}_n^{(4)}{\cdot}{\cal E}_m^{(4)} \right)
\end{split}
\ee
As expected the factorization of the 4-HES scattering is consistent with the structure of the covariant version of the three Reggeon interaction. The $t$-channel factorization can be derived, as in the case of the Veneziano amplitude, by using the symmetry of the amplitude that is manifestly preserved by the dressing of the HES which is reflected by the symmetries of the Jacobi polynomials. As expected the $t$-channel factorization gives 
\be
{\cal A}_{gen}^{4HES}(s,t)\Big|_{P_{N}^{2}\sim N-1}=g_{o}^{2}{\cal A}_{gen}^{2HES + N}(p_{2},p_{3},P_{N}) {{\cal Z}_{N}(\ell_{s}^{2}s{+}2)\over {-}\ell_{s}^{2}P_{N}^{2}{-}1{+}N } {\cal A}_{gen}^{2HES + N}(p_{1},p_{4},-P_{N})
\ee

\section{Shapiro-Virasoro amplitude of four arbitrarily excited strings}\label{SVsec}
In this section we extend the computation of sec.\ref{VenHES} to the case of tree level scattering of four closed bosonic arbitrarily excited string states. Starting from the coherent vertex operators in (\ref{closedV}), the insertion of four punctures on the sphere is given by 
\be\label{C4}
{\cal M}_{gen}^{HES}(s,t,u)=g_{c}^{4}\ell_{s}^{4}C_{S_{2}}\int_{S_{2}} \prod_{\ell=1}^{4}d^{2}z_{\ell}\,\Bigg\langle \prod_{\ell=1}^{4}V_{{\cal C}}(p_{\ell};z_{\ell},\overline z_{\ell})\Bigg\rangle_{S_{2}}
\ee
\begin{figure}[h!]
\centering
\includegraphics[scale=0.5]{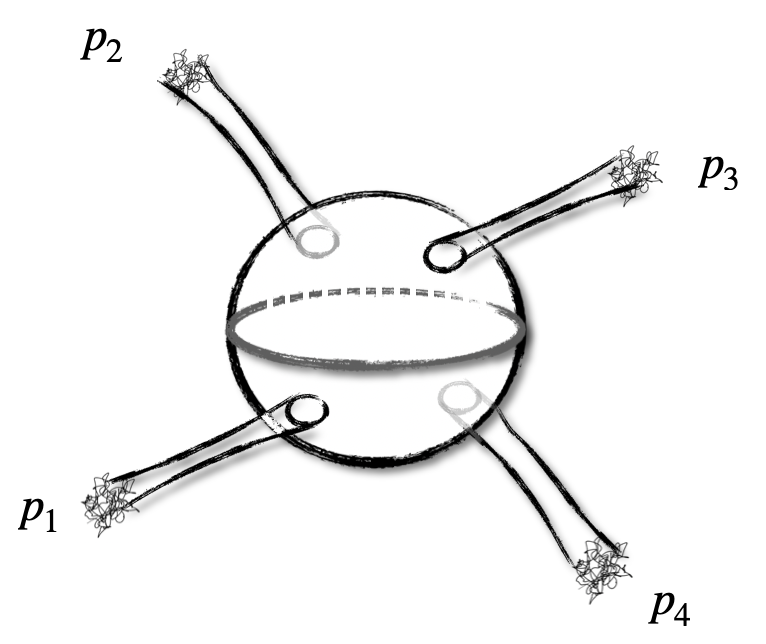}
\caption{Sphere amplitude with four HES punctures.}
\end{figure}
since the correlation function on the sphere can be separated into holomorphic and anti-holomorphic part as 
\be
\Bigg\langle \prod_{\ell=1}^{4}V_{{\cal C}}(p_{\ell};z_{\ell},\overline z_{\ell})\Bigg\rangle_{S_{2}}=\Bigg\langle \prod_{\ell=1}^{4}V_{{\cal O}}(p_{\ell};z_{\ell})\Bigg\rangle_{D_{2}}\Bigg\langle \prod_{\ell=1}^{4}\overline V_{{\cal O}}(p_{\ell};\overline z_{\ell})\Bigg\rangle_{D_{2}}
\ee
immediately one gets the same contractions present in the exponential of (\ref{intpreVen}) for both correlation functions, with the closed string version of the Koba-Nielsen factor (\ref{KN}) where $\ell_{s}\rightarrow \ell_{s}/2$. Taking into account the redundancy of $SL(2,C)$, repeating the same steps of sec.\ref{VenHES} and app.\ref{app1} one gets 
\be\label{intampc}
\begin{split}
{\cal M}_{gen}^{4HES}(s,t,u)&=g_{c}^{2}\int_{S_{2}}d^{2}z\, z^{-{\ell_{s}^{2}s\over 4}{-}2}(1{-}z)^{-{\ell_{s}^{2}t\over 4}{-}2}\overline z^{-{\ell_{s}^{2}s\over 4}{-}2}(1{-}\overline z)^{-{\ell_{s}^{2}t\over 4}{-}2}\\
&\exp{\left(\sum_{\ell}\sum_{n_{\ell}}V^{(\ell)}_{n_{\ell}}(z)+{1\over 2}\sum_{\ell}\sum_{n_{\ell},m_{\ell}}W^{(\ell)}_{n_{\ell},m_{\ell}}(z)+{1\over 2}\sum_{v,f}\sum_{n_{v},m_{f}} I_{n_{v},m_{f}}^{(v,f)}(z)\right)}\\
&\exp{\left(\sum_{\ell}\sum_{n_{\ell}}\overline V^{(\ell)}_{n_{\ell}}(\overline z)+{1\over 2}\sum_{\ell}\sum_{n_{\ell},m_{\ell}}\overline W^{(\ell)}_{n_{\ell},m_{\ell}}(\overline z)+{1\over 2}\sum_{v,f}\sum_{n_{v},m_{f}} \overline I_{n_{v},m_{f}}^{(v,f)}(\overline z)\right)}
\end{split}
\ee
Now using the analogue of (\ref{formVen}) with the parametrization (\ref{idVen}) for holomorphic and anti-holomorphic parts
\be\label{idSV}
z^{k}\rightarrow \partial^{k}_{\beta_{s}}\,,\quad (1{-}z)^{k}\rightarrow \partial^{k}_{\beta_{t}}\quad; \quad \overline z^{k}\rightarrow \partial^{k}_{\overline\beta_{s}}\,,\quad (1{-}\overline z)^{k}\rightarrow \partial^{k}_{\overline \beta_{t}}
\ee
the amplitude can then be represented as
\be
\begin{split}
&{\cal M}_{gen}^{4HES}(s,t,u)=g_{c}^{2}e^{{\cal K}\left(\{\zeta^{(\ell)}_{n}\}; \partial_{\beta_{s}},\partial_{\beta_{t}}\right)}e^{\overline{\cal K}\left(\{\overline\zeta^{(\ell)}_{n}\}; \partial_{\overline\beta_{s}},\partial_{\overline\beta_{t}}\right)}\\
&\int_{S_{2}}d^{2}z\, z^{-{\ell_{s}^{2}s\over 4}{-}2}(1{-}z)^{-{\ell_{s}^{2}t\over 4}{-}2}e^{\beta_{s}z}e^{\beta_{t}(1{-}z)} \overline z^{-{\ell_{s}^{2}s\over 4}{-}2}(1{-}\overline z)^{-{\ell_{s}^{2}t\over 4}{-}2}e^{\overline\beta_{s}\overline z}e^{\overline\beta_{t}(1{-}\overline z)}\Big|_{\beta_{s,t},\overline\beta_{s,t}=0}
\end{split}
\ee
where ${\cal K}$ and $\overline {\cal K}$ are the extension of (\ref{Kfunc}). Expanding the exponentials and using the general formula for the integration on the sphere
\be
\int_{S_{2}}d^{2}z \,z^{a}(1{-}z)^{b}\overline z^{\overline a}(1{-}\overline z)^{\overline b}={\Gamma(1{+}a)\Gamma(1{+}b)\over \Gamma(2{+}a{+}b)}{\Gamma(-1{-}\overline a{-}\overline b)\over \Gamma(-\overline a)\Gamma(-\overline b)}
\ee
the complete expression of the scattering amplitude of four arbitrarily excited states in the closed string case is given by
\be\label{SVamp}
\begin{split}
{\cal M}_{gen}^{4HES}(s,t,u)&=g_{c}^{2}e^{{\cal K}\left(\{\zeta^{(\ell)}_{n}\}; \partial_{\beta_{s}},\partial_{\beta_{t}}\right)}e^{\overline{\cal K}\left(\{\overline\zeta^{(\ell)}_{n}\}; \partial_{\overline\beta_{s}},\partial_{\overline\beta_{t}}\right)}\sum_{a_{s},a_{t}}{\beta_{s}^{a_{s}}\over a_{s}!}{\beta_{t}^{a_{t}}\over a_{t}!}\sum_{\overline a_{s},\overline a_{t}}{\overline \beta_{s}^{\overline a_{s}}\over \overline a_{s}!}{\overline \beta_{t}^{\overline a_{t}}\over \overline a_{t}!}\\
&{\Gamma\left(-{\ell_{s}^{2}s\over4}{-}1{+}a_{s}\right)\Gamma\left(-{\ell_{s}^{2}t\over4}{-}1{+}a_{t}\right)\Gamma\left(-{\ell_{s}^{2}u\over4}{-}1{+}\sum_{\ell}N_{\ell}{-}\overline a_{s}{-}\overline a_{t}\right)\over \Gamma\left({\ell_{s}^{2}u\over4}{+}2{-}\sum_{\ell}N_{\ell}{+}a_{s}{+}a_{t}\right)\Gamma\left({\ell_{s}^{2}s\over4}{+}2{-}\overline a_{s}\right)\Gamma\left({\ell_{s}^{2}t\over4}{+}2{-}\overline a_{t}\right) }\Big|_{\beta_{s,t},\overline\beta_{s,t}=0}
\end{split}
\ee
where it was used the relation 
\be
{\ell_{s}^{2}s\over 4}+{\ell_{s}^{2}t\over 4}+{\ell_{s}^{2}u\over 4}=\sum_{\ell=1}^{4}N_{\ell}-4
\ee
The result (\ref{SVamp}) can be massaged in a more compact expression, where still one can separate the poles structure from the contributions of external states, leading to
\be
{\cal M}_{gen}^{4HES}(s,t,u)={\cal M}_{SV}(s,t,u)e^{{\cal K}\left(\{\zeta^{(\ell)}_{n}\}; \partial_{\beta_{s}},\partial_{\beta_{t}}\right)}\Phi_{\beta_{s},\beta_{t}}(s,t)e^{\overline{\cal K}\left(\{\overline\zeta^{(\ell)}_{n}\}; \partial_{\overline\beta_{s}},\partial_{\overline\beta_{t}}\right)}\overline\Phi_{\overline\beta_{s},\overline\beta_{t}}(s,t)
\ee
As one can see, the first part is the Shapiro-Virasoro amplitude 
\be
{\cal M}_{SV}(s,t,u)=g_{c}^{2}{\Gamma\left(-{\ell_{s}^{2}s\over 4}{-}1\right)\Gamma\left(-{\ell_{s}^{2}t\over 4}{-}1\right)\Gamma\left(-{\ell_{s}^{2}u\over 4}{-}1{+}\sum_{\ell}N_{\ell}\right)\over \Gamma\left({\ell_{s}^{2}s\over 4}{+}2 \right)\Gamma\left({\ell_{s}^{2}t\over 4}{+}2 \right)\Gamma\left({\ell_{s}^{2}u\over 4}{+}2{-}\sum_{\ell}N_{\ell} \right)}
\ee
while the dressing factors contain the external state information, as in the open string case with (\ref{Kfunc}) and (\ref{ffunc}). In this compact representation is very easy to see the natural generalization of the KLT \cite{Kawai:1985xq} relations applied to the scattering of four arbitrarily excited string states
\be
{\cal M}_{gen}^{4HES}(s,t,u)=\sin{\pi\ell_{s}^{2}t\over 4}{\cal A}^{4HES}_{gen}(s,t){\cal A}^{4HES}_{gen}(t,u)
\ee
Since quite recently the KLT relations were extended to one loop amplitudes \cite{Stieberger:2023nol}, it would be absolutely interesting but also extremely challenging to repeat a similar analysis in the case of arbitrarily excited string states.

The scattering amplitude with any given state can be extracted by applying the following derivative projections on the holomorphic and anti-holomorphic sectors respectively 
\be
{\cal D}^{N_{\ell}}_{\{g_{n_{\ell}}\}}(\{\epsilon_{a,n}^{(\ell)}\})=\prod_{n_{\ell}=1}^{N_{\ell}}{1\over \sqrt{n_{\ell}^{g_{n_{\ell}}}g_{n_{\ell}}!}}\prod_{a=1}^{g_{n_{\ell}}}\epsilon_{a,n}^{(\ell)}{\cdot}{d\over d\zeta_{n_{\ell}}^{(\ell)}}
\ee
\be
\overline{\cal D}^{\overline N_{\ell}}_{\{\overline g_{n_{\ell}}\}}(\{\overline\epsilon_{a,n}^{(\ell)}\})=\prod_{n_{\ell}=1}^{\overline N_{\ell}}{1\over \sqrt{n_{\ell}^{\overline g_{n_{\ell}}}\overline g_{n_{\ell}}!}}\prod_{a=1}^{\overline g_{n_{\ell}}}\overline\epsilon_{a,n}^{(\ell)}{\cdot}{d\over d\overline\zeta_{n_{\ell}}^{(\ell)}}
\ee
which is describing the closed string degeneracy as the square of the open string degeneracy where also the level matching condition must be imposed 
\be
\sum_{n=1}^{N}ng_{n}=N=\overline N=\sum_{n=1}^{\overline N}n\overline g_{n}
\ee
In this representation the scattering amplitude of four closed string states is given by 
\be\label{SVproj}
{\cal M}_{N_{1},N_{2},N_{3},N_{4}}^{4HES}(s,t)={\cal D}^{N_{\ell}}_{\{g_{n_{\ell}}\}}(\{\epsilon_{a,n}^{(\ell)}\})\overline{\cal D}^{\overline N_{\ell}}_{\{\overline g_{n_{\ell}}\}}(\{\overline\epsilon_{a,n}^{(\ell)}\}){\cal M}_{gen}^{HES}(s,t,u)\Big|_{\{\zeta_{n_{\ell}}\}=0}
\ee

\section{Conclusion and future directions}\label{concl}
In the present paper it was described the extension of the Veneziano and Shapiro-Virasoro amplitudes to the case of arbitrarily excited states in bosonic string. We analyzed the factorization properties of the extended Veneziano amplitude, and as a result we found a compact expression in terms of the covariant version of the three string interaction that we matched with the result of the old fashioned computation.
Relying on the compact version of the Shapiro-Virasoro amplitude, we also found the extended version of the KLT relations applied to arbitrarily excited states.
The advantage of having closed form polynomials, representing the role of arbitrarily excited external states, is linked to the possibility of further explore asymptotic properties of string interactions parametrizing for examples large excitation level contributions or large degeneracy contributions.

Our results are in the context of bosonic string, but we hope to implement a similar construction also in superstring theory, where from one side a superstring vertex operator is known \cite{Aldi:2019osr} and from the other side the superstring version of the three reggeon vertex is also known \cite{Hornfeck:1987wt}.

Regarding the compact embedding of the external arbitrarily excited states contributions in the scattering amplitude, it is natural to comment about the $N$-point scattering generalization of such states. In fact following the cyclic representation in (\ref{intpreVen}), one can extend the Koba-Nielsen contribution to $N$ punctures as well as the number of Wick contractions of external states. But in oder to construct manifestly invariant terms, under conformal invariance, one has to find an algorithmic way to implement the M\"obius transformations for $N$ punctures, which is a technical point that we hope to address in future studies.

Another important aspect, we would like to further explore, is a possible connection with off-shell regularization of scattering amplitudes \cite{Schwarz:1973gxz}-\cite{Bluhm:1988ew}. Intuitively the picture provided by scattering coherent states seems to be intimately connected with some off-shell parametrization of string amplitudes, but at this stage we are not able to make any precise statement. Nevertheless, since the three string interaction vertex was generalized to non flat backgrounds \cite{Schwarz:2003zf}\cite{DiVecchia:2003yp}, it would be very interesting to explore if, a construction similar to our extension of the Veneziano and Shapiro-Virasoro amplitudes, can be done in the context of AdS/CFT correspondence \cite{Aharony:1999ti} in order to extract additional information about the correlation functions of the boundary theory \cite{Bianchi:2002rw}-\cite{Minahan:2012fh}.

In the context of gravitational interactions, it would be also very interesting to study string tidal deformations in high energy interactions of four arbitrarily excited states, providing a quantitative analysis of the scattering properties as suggest in \cite{Giddings:2007bw}\cite{Veneziano:2004er}. In fact due to the chaotic behavior of HES amplitudes, it is not clear if there is a chaos driven counterpart in the characterization of black hole formation in transplankian or ultra-high energy processes.

\section*{Acknowledgments}
I would like to thank Paolo Di Vecchia, Arkady Tseytlin, Giancarlo Rossi, Elias Kiritsis, Vasilis Niarchos, Bo Sundborg and Massimo Bianchi for useful comments.
Finally I want to thank the Nordic Institute for Theoretical Physics (NORDITA) for their hospitality during a period when part of the research work has been conducted.
This research is partially supported by the European  MSCA grant HORIZON-MSCA-2022-PF-01-01 "BlackHoleChaos" No.101105116 and by the H.F.R.I call 
“Basic research Financing (Horizontal support of all Sciences)” under 
the National Recovery and Resilience Plan “Greece 2.0” funded by the 
European Union – NextGenerationEU (H.F.R.I. Project Number: 15384.).

\begin{appendix}
\section{Wick contractions of the generating amplitude and $SL(2,\mathit{R})$ covariance}\label{app1}
In the present appendix we want to make manifest the $SL(2,\mathit{R})$ covariance of all the Wick contractions between the arbitrarily excited string vertex operators. In particular we will show how the M\"obius transformations generate closed form polynomial expressions that we identify as combinations of Jacobi polynomials.

The closed string case is obtained as the repetition of the manipulations of the open string terms, since the $SL(2,\mathit{C})$ covariance works separately in the holomorphic and anti-holomorphic part of the correlation functions.
\subsection{Contraction with single polarization}
We want to manipulate the contractions
\be
-\left({z_{\ell,\ell+1}z_{\ell,\ell-1}\over z_{\ell+1,\ell-1}} \right)^{n} {\cal V}_{n}^{(\ell)}(z_{j\ell})
\ee
in order to show that the $SL(2,\mathit{R})$ manifestly covariant term is given by 
\be
-\left({z_{\ell,\ell+1}z_{\ell,\ell-1}\over z_{\ell+1,\ell-1}} \right)^{n} {\cal V}_{n}^{(\ell)}(z_{j\ell})\Rightarrow V^{(\ell)}_{n}(z)
\ee
where the final polynomial can be casted in the following form
\be\label{expr2}
V^{(\ell)}_{n}(z)=(-)^{n}\sqrt{2}\ell_{s}\zeta_{n}^{(\ell)}{\cdot}p_{\ell+1}P_{n-1}^{\hspace{0 mm}^{\hspace{0 mm}^{(\alpha_{\ell}^{(n)},\beta_{\ell}^{(n)})}}}\hspace{-9mm}(1{-}2 R_{\ell})+(-)^{n}R_{\ell}\sqrt{2}\ell_{s}\zeta_{n}^{(\ell)}{\cdot}p_{\ell+2} P_{n-1}^{\hspace{0 mm}^{\hspace{0 mm}^{(\alpha_{\ell}^{(n)}{+}1,\beta_{\ell}^{(n)})}}}\hspace{-12mm}(1{-}2 R_{\ell})
\ee
with
\be
\alpha_{\ell}^{(n)}=-n-2\ell_{s}^{2}nq_{\ell}{\cdot}p_{\ell+1}\,,\quad \beta_{\ell}^{(n)}=-n-2\ell_{s}^{2}nq_{\ell}{\cdot}p_{\ell-1}\,,\quad R_{\ell}={z_{\ell+1,\ell}\over z_{\ell+1,\ell-1}}{z_{\ell+2,\ell-1}\over z_{\ell+2,\ell}}
\ee
In order to prove the result one has to start from 
\be\label{this}
-\left({z_{\ell,\ell+1}z_{\ell,\ell-1}\over z_{\ell+1,\ell-1}} \right)^{n}\sum_{k=1}^{n}\left(\sum_{j\ne \ell}{\sqrt{2}\ell_{s}\zeta_{n}^{(\ell)}{\cdot}p_{j}\over z_{j\ell}^{k}}\right){\cal Z}_{n-k}\left(n\sum_{j\ne\ell}{2\ell_{s}^{2}q_{\ell}{\cdot}p_{j}\over z^{f}_{j\ell}} \right)
\ee
without loss of generality we can set for the moment $\sqrt{2}\ell_{s}=1$, and from the explicit expression of the cycle index polynomial 
\be\label{cpoly}
{\cal Z}_{n-k}\left(n\sum_{j\ne\ell}{q_{\ell}{\cdot}p_{j}\over z^{f}_{j\ell}} \right)=\oint {d\omega\over \omega^{n{-}k{+}1}}\prod_{j\ne \ell}\left( 1{-}{\omega\over z_{j\ell}}\right)^{-nq_{\ell}{\cdot}p_{j}}
\ee
one can write the expression (\ref{this}) as
\be
-\left({z_{\ell,\ell+1}z_{\ell,\ell-1}\over z_{\ell+1,\ell-1}} \right)^{n}\sum_{k=1}^{n}\left(\sum_{j\ne \ell}{\zeta_{n}^{(\ell)}{\cdot}p_{j}\over z_{j\ell}^{k}}\right)\oint {d\omega\over \omega^{n{-}k{+}1}}\prod_{j\ne \ell}\left( 1{-}{\omega\over z_{j\ell}}\right)^{-nq_{\ell}{\cdot}p_{j}}
\ee 
and after computing the sum 
\be
\sum_{k=1}^{n} \left({\omega\over z_{j\ell}}\right)^{k}={\omega\over z_{j\ell}-\omega}
\ee
one ends with
\be
\left({z_{\ell,\ell+1}z_{\ell,\ell-1}\over z_{\ell+1,\ell-1}} \right)^{n}\oint {d\omega\over \omega^{n}}\left(\sum_{j\ne \ell}{\zeta_{n}^{(\ell)}{\cdot}p_{j}\over \omega{-}z_{j\ell}}\right)\prod_{j\ne \ell}\left( 1{-}{\omega\over z_{j\ell}}\right)^{-nq_{\ell}{\cdot}p_{j}}
\ee 
From momentum conservation we can write 
\be
\sum_{j\ne \ell}{\zeta_{n}^{(\ell)}{\cdot}p_{j}\over \omega{-}z_{j\ell}}={z_{\ell+1,\ell-1}\over z_{\ell+1,\ell}z_{\ell-1,\ell}}{\zeta_{n}^{(\ell)}{\cdot}p_{\ell+1}\over \left(1{-}{\omega\over z_{\ell+1,\ell}}\right) \left(1{-}{\omega\over z_{\ell-1,\ell}}\right)}+{z_{\ell+2,\ell-1}\over z_{\ell+2,\ell}z_{\ell-1,\ell}}{\zeta_{n}^{(\ell)}{\cdot}p_{\ell+2}\over \left(1{-}{\omega\over z_{\ell+2,\ell}}\right) \left(1{-}{\omega\over z_{\ell-1,\ell}}\right)}
\ee
splitting the expression in two main contributions
\be
\begin{split}
&\left({z_{\ell,\ell+1}z_{\ell,\ell-1}\over z_{\ell+1,\ell-1}} \right)^{n-1}\zeta_{n}^{(\ell)}{\cdot}p_{\ell+1}\oint {d\omega\over \omega^{n}}\left( 1{-}{\omega\over z_{\ell+2,\ell}}\right)\prod_{j\ne \ell}\left( 1{-}{\omega\over z_{j\ell}}\right)^{-nq_{\ell}{\cdot}p_{j}-1}\\
&+\left({z_{\ell,\ell+1}z_{\ell,\ell-1}\over z_{\ell+1,\ell-1}} \right)^{n-1}{z_{\ell+1,\ell}\over z_{\ell+1,\ell-1}}{z_{\ell+2,\ell-1}\over z_{\ell+2,\ell}}
\zeta_{n}^{(\ell)}{\cdot}p_{\ell+2}\oint {d\omega\over \omega^{n}}\left( 1{-}{\omega\over z_{\ell+1,\ell}}\right)\prod_{j\ne \ell}\left( 1{-}{\omega\over z_{j\ell}}\right)^{-nq_{\ell}{\cdot}p_{j}-1}
\end{split}
\ee
now using the transformation
\be
\omega={z_{\ell-1,\ell}\over \rho}\,,\quad d\omega=-z_{\ell-1.\ell}{d\rho\over \rho}
\ee
from the first term one gets
\be
\begin{split}
-\left({z_{\ell,\ell+1}z_{\ell,\ell-1}\over z_{\ell+1,\ell-1}} \right)^{n-1}\zeta_{n}^{(\ell)}{\cdot}p_{\ell+1}&\oint d\rho{\rho^{n-2}\over z_{\ell-1,\ell}^{n-1} }\left( 1{-}{1\over \rho}\right)^{-nq_{\ell}{\cdot}p_{\ell-1}-1}\\
&\left( 1{-}{z_{\ell-1,\ell}\over z_{\ell+2,\ell}\rho}\right)^{-nq_{\ell}{\cdot}p_{\ell+2}}\left( 1{-}{z_{\ell-1,\ell}\over z_{\ell+1,\ell}\rho}\right)^{-nq_{\ell}{\cdot}p_{\ell+1}-1}
\end{split}
\ee
while from the second 
\be 
\begin{split}
-\left({z_{\ell,\ell+1}z_{\ell,\ell-1}\over z_{\ell+1,\ell-1}} \right)^{n-1}{z_{\ell+1,\ell}\over z_{\ell+1,\ell-1}}{z_{\ell+2,\ell-1}\over z_{\ell+2,\ell}}
&\zeta_{n}^{(\ell)}{\cdot}p_{\ell+2}\oint d\rho{\rho^{n-2}\over z_{\ell-1,\ell}^{n-1} }\left( 1{-}{1\over \rho}\right)^{-nq_{\ell}{\cdot}p_{\ell-1}-1}\\
&\left( 1{-}{z_{\ell-1,\ell}\over z_{\ell+2,\ell}\rho}\right)^{-nq_{\ell}{\cdot}p_{\ell+2}-1}\left( 1{-}{z_{\ell-1,\ell}\over z_{\ell+1,\ell}\rho}\right)^{-nq_{\ell}{\cdot}p_{\ell+1}}
\end{split}
\ee 
still from the momentum conservation
\be\label{quella}
\rho^{n}=\rho^{-nq_{\ell}{\cdot}p_{\ell-1}}\rho^{-nq_{\ell}{\cdot}p_{\ell+1}}\rho^{-nq_{\ell}{\cdot}p_{\ell+2}}
\ee 
 the previous expressions can be written as
 \be
\begin{split}
(-)^{n}\left({z_{\ell,\ell+1}\over z_{\ell+1,\ell-1}} \right)^{n-1}\zeta_{n}^{(\ell)}{\cdot}p_{\ell+1}&\oint d\rho \left( \rho{-}1\right)^{-nq_{\ell}{\cdot}p_{\ell-1}-1}\\
&\left( \rho{-}{z_{\ell-1,\ell}\over z_{\ell+2,\ell}}\right)^{-nq_{\ell}{\cdot}p_{\ell+2}}\left( \rho{-}{z_{\ell-1,\ell}\over z_{\ell+1,\ell}}\right)^{-nq_{\ell}{\cdot}p_{\ell+1}-1}
\end{split}
\ee
and
\be 
\begin{split}
(-)^{n}\left({z_{\ell,\ell+1}\over z_{\ell+1,\ell-1}} \right)^{n-1}{z_{\ell+1,\ell}\over z_{\ell+1,\ell-1}}{z_{\ell+2,\ell-1}\over z_{\ell+2,\ell}}
&\zeta_{n}^{(\ell)}{\cdot}p_{\ell+2}\oint d\rho\left( \rho{-}1\right)^{-nq_{\ell}{\cdot}p_{\ell-1}-1}\\
&\left( \rho{-}{z_{\ell-1,\ell}\over z_{\ell+2,\ell}}\right)^{-nq_{\ell}{\cdot}p_{\ell+2}-1}\left( 1{-}{z_{\ell-1,\ell}\over z_{\ell+1,\ell}}\right)^{-nq_{\ell}{\cdot}p_{\ell+1}}
\end{split}
\ee
Finally using the transformation
\be
\rho=1+{1\over \eta}{z_{\ell+1,\ell-1}\over z_{\ell+1,\ell}}\,,\quad d\rho=-{z_{\ell+1,\ell-1}\over z_{\ell+1,\ell}}{d\eta\over \eta^{2}}
\ee
where
\be
\rho-1={1\over \eta}{z_{\ell+1,\ell-1}\over z_{\ell+1,\ell}}\,,\quad \rho-{z_{\ell-1,\ell}\over z_{\ell+1,\ell}}={1\over \eta}{z_{\ell+1,\ell-1}\over z_{\ell+1,\ell}}\left(1+\eta\right)
\ee
\be
\rho{-}{z_{\ell-1,\ell}\over z_{\ell+2,\ell}}={1\over \eta}{z_{\ell+1,\ell-1}\over z_{\ell+1,\ell}}\left(1+\eta {z_{\ell+1,\ell}z_{\ell+2,\ell-1}\over z_{\ell+1,\ell-1}z_{\ell+2,\ell}}\right)
\ee
and using the momentum conservation (\ref{quella}) for the $\eta$ variable one gets
\be\label{exprVl}
\begin{split}
&\zeta_{n}^{(\ell)}{\cdot}p_{\ell+1}\oint {d\eta\over \eta^{n}} (1{+}\eta)^{-nq_{\ell}{\cdot}p_{\ell+1}-1}\left(1{+}\eta {z_{\ell+1,\ell}\over z_{\ell+1,\ell-1}}{z_{\ell+2,\ell-1}\over z_{\ell+2,\ell}}\right)^{-nq_{\ell}{\cdot}p_{\ell+2}}+\\
&{z_{\ell+1,\ell}\over z_{\ell+1,\ell-1}}{z_{\ell+2,\ell-1}\over z_{\ell+2,\ell}}\zeta_{n}^{(\ell)}{\cdot}p_{\ell+2}\oint {d\eta\over \eta^{n}} (1{+}\eta)^{-nq_{\ell}{\cdot}p_{\ell+1}}\left(1{+}\eta {z_{\ell+1,\ell}\over z_{\ell+1,\ell-1}}{z_{\ell+2,\ell-1}\over z_{\ell+2,\ell}}\right)^{-nq_{\ell}{\cdot}p_{\ell+2}-1}
\end{split}
\ee
These expressions can be represented in terms of the cycle index polynomial, in fact taking the identity 
\be
\begin{split}
\oint {d\sigma'\over {\sigma'}^{n}}(1+\sigma')^{-a}(1+R\sigma')^{-b}\Big|_{\sigma'{\rightarrow} -\sigma}&=(-)^{n+1}\oint {d\sigma\over {\sigma}^{n}}(1-\sigma)^{-a}(1-R\sigma)^{-b}\\
&=(-)^{n+1}\sum_{k=1}^{n}\oint {d\sigma\over {\sigma}^{n-k+1}}(1-\sigma)^{-a+1}(1-R\sigma)^{-b}\\
\end{split}
\ee
where in the last step it was used 
\be
\sigma=(1{-}\sigma)\sum_{k=1}^{\infty}\sigma^{k}
\ee
one can see from (\ref{cpoly}) that 
\be
\sum_{k=1}^{\infty}\oint {d\sigma\over {\sigma}^{n-k+1}}(1-\sigma)^{-a+1}(1-R\sigma)^{-b}=\sum_{k=1}^{n}Z_{n-k}\Big(a-1 +R^{r} b \Big)
\ee
An explicit expression for this kind of polynomials is provided by the following properties of the cycle index polynomial
\be\label{properties}
{\cal Z}_{n}(a_{\ell}+b_{\ell})=\sum_{k=0}^{n}{\cal Z}_{n-k}(a_{\ell}){\cal Z}_{k}(b_{\ell})\,,\quad {\cal Z}_{n}(z^{\ell}a_{\ell})=z^{n}{\cal Z}_{n}(a_{\ell})\,,\quad {\cal Z}_{n}(z)={(z)_{n}\over n!}
\ee
leading to the identity 
\be
\sum_{k=1}^{\infty}Z_{n-k}\Big(a-1 +R^{r} b \Big)= {(a)_{n-1}\over \Gamma(n)} \, \pFq{2}{1}{ b, 1{-}n}{ 1{-}n{-}(a{-}1)}{ \,R}
\ee 
and finally, introducing the Jacobi polynomial $P_{N}^{(\alpha,\beta)}(x)$
\be
P_{N}^{(\alpha,\beta)}(x)=\sum_{r=0}^{N}\begin{pmatrix}N+\alpha\\N-r\end{pmatrix}\begin{pmatrix}N+\beta\\r\end{pmatrix}\left({x-1\over 2}\right)^{r}\left({x+1\over 2}\right)^{N-r}
\ee
 and using the relation
\be
{N!\over (\alpha+1)_{N}} P_{N}^{(\alpha,\beta)}(1-2 R)=\pFq{2}{1}{ \alpha+1+\beta+N, {-}N}{ \alpha+1}{ \,R}
\ee
with the identifications
\be
N=n-1\,,\quad \alpha=-n-(a-1)\,,\quad \beta=b+a-1
\ee
 one can write
\be\label{Jpolyhyp}
\begin{split}
 {(a)_{n-1}\over \Gamma(n)} \, \pFq{2}{1}{ b, 1{-}n}{ 1{-}n{-}(a{-}1)}{ \,z}&={(a)_{n-1}\over (2{-}n{-}a)_{n-1}}P_{n-1}^{(-n-(a-1),b+a-1)}(1{-}2 R)\\
 &=(-)^{n+1}P_{n-1}^{(-n-(a-1),b+a-1)}(1{-}2 R)
 \end{split}
\ee
Applying this procedure to (\ref{exprVl}), the first term is identified by $a=1+nq_{\ell}{\cdot}p_{\ell+1}$ and $b=nq_{\ell}{\cdot}p_{\ell+2}$ while for the second term one has $a=nq_{\ell}{\cdot}p_{\ell+1}$ and $b=1+nq_{\ell}{\cdot}p_{\ell+2}$ and the result, where we have reintroduced $\ell_{s}$, is given by 
\be\label{expr2}
V^{(\ell)}_{n}(z)=\sqrt{2}\ell_{s}\zeta_{n}^{(\ell)}{\cdot}p_{\ell+1}P_{n-1}^{\hspace{0 mm}^{\hspace{0 mm}^{(\alpha_{\ell}^{(n)},\beta_{\ell}^{(n)})}}}\hspace{-9mm}(1{-}2 R_{\ell})+R_{\ell}\sqrt{2}\ell_{s}\zeta_{n}^{(\ell)}{\cdot}p_{\ell+2} P_{n-1}^{\hspace{0 mm}^{\hspace{0 mm}^{(\alpha_{\ell}^{(n)}{+}1,\beta_{\ell}^{(n)})}}}\hspace{-12mm}(1{-}2 R_{\ell})
\ee
where
\be\label{coeffJac}
\alpha_{\ell}^{(n)}=-n-2\ell_{s}^{2}nq_{\ell}{\cdot}p_{\ell+1}\,,\quad \beta_{\ell}^{(n)}=-n-2\ell_{s}^{2}nq_{\ell}{\cdot}p_{\ell-1}\,,\quad R_{\ell}={z_{\ell+1,\ell}\over z_{\ell+1,\ell-1}}{z_{\ell+2,\ell-1}\over z_{\ell+2,\ell}}
\ee
\subsection{Contractions with two polarizations}
In what follows we will manipulate the contractions
\be
 \left({z_{\ell,\ell+1}z_{\ell,\ell-1}\over z_{\ell+1,\ell-1}} \right)^{n_{\ell}+m_{\ell}}\hspace{-4mm} {\cal W}^{(\ell)}_{n_{\ell},m_{\ell}}(z_{j\ell})
\ee
and as a result of the $SL(2,\mathit{R})$ covariance, we will show  
\be
{\zeta^{(\ell)}_{n_{\ell}}{\cdot}\zeta^{(\ell)}_{m_{\ell}}} \left({z_{\ell,\ell+1}z_{\ell,\ell-1}\over z_{\ell+1,\ell-1}} \right)^{n_{\ell}+m_{\ell}}\hspace{-4mm} {\cal W}^{(\ell)}_{n_{\ell},m_{\ell}}(z_{j\ell})\Rightarrow W^{(\ell)}_{n_{\ell},m_{\ell}}(z)
\ee
where
\be\label{finW}
W^{(\ell)}_{n_{\ell},m_{\ell}}(z)={\zeta^{(\ell)}_{n_{\ell}}{\cdot}\zeta^{(\ell)}_{m_{\ell}}}\sum_{r=1}^{m_{\ell}}r\,P_{n_{\ell}+r}^{\hspace{0 mm}^{\hspace{0 mm}^{(\alpha_{\ell}^{(n_{\ell})}-r,\beta_{\ell}^{(n_{\ell})}-1)}}}\hspace{-16.5mm}(1{-}2 R_{\ell})\,\,P_{m_{\ell}-r}^{\hspace{0 mm}^{\hspace{0 mm}^{(\alpha_{\ell}^{(m_{\ell})}+r,\beta_{\ell}^{(m_{\ell})}-1)}}}\hspace{-17mm}(1{-}2 R_{\ell})
\ee
with
\be\label{coeffJac}
\alpha_{\ell}^{(n)}=-n-2\ell_{s}^{2}nq_{\ell}{\cdot}p_{\ell+1}\,,\quad \beta_{\ell}^{(n)}=-n-2\ell_{s}^{2}nq_{\ell}{\cdot}p_{\ell-1}\,,\quad R_{\ell}={z_{\ell+1,\ell}\over z_{\ell+1,\ell-1}}{z_{\ell+2,\ell-1}\over z_{\ell+2,\ell}}
\ee
Let us consider the Wick contraction
\be\label{conW}
\begin{split}
&\left({z_{\ell,\ell+1}z_{\ell,\ell-1}\over z_{\ell+1,\ell-1}} \right)^{n_{\ell}+m_{\ell}}{\cal W}^{(\ell)}_{n_{\ell},m_{\ell}}(z_{j\ell})=\\
&\left({z_{\ell,\ell+1}z_{\ell,\ell-1}\over z_{\ell+1,\ell-1}} \right)^{n_{\ell}+m_{\ell}} \sum_{r=1}^{m_{\ell}}r\,{\cal Z}_{n_{\ell}{+}r}\left(n_{\ell}\sum_{j\ne\ell}{2\ell_{s}^{2}q_{\ell}{\cdot}p_{j}\over z^{f}_{j\ell}} \right){\cal Z}_{m_{\ell}{-}r}\left(m_{\ell}\sum_{j\ne\ell}{2\ell_{s}^{2}q_{\ell}{\cdot}p_{j}\over z^{f}_{j\ell}} \right)
\end{split}
\ee
setting for simplicity $\sqrt{2}\ell_{s}=1$, with the help of (\ref{cpoly}) one can compute the sum 
\be\label{sum2pol}
\sum_{r=1}^{\infty}r \left({\omega_{2}\over \omega_{1}}\right)^{r}={\omega_{1}\omega_{2}\over (\omega_{1}{-}\omega_{2})^{2}}
\ee
that used in (\ref{conW}) yields 
 \be
\left( z_{\ell,\ell+1}z_{\ell,\ell-1}\over z_{\ell+1,\ell-1} \right)^{n_{\ell}+m_{\ell}} \oint{d\omega_{1}\over \omega_{1}^{n_{\ell}}}\oint{d\omega_{2}\over \omega_{2}^{m_{\ell}}}{1\over (\omega_{1}-\omega_{2})^{2}}\prod_{j\ne\ell}\left( 1{-}{\omega_{1}\over z_{j\ell}}\right)^{-n_{\ell}q_{\ell}{\cdot}p_{j}}\left( 1{-}{\omega_{2}\over z_{j\ell}}\right)^{-m_{\ell}q_{\ell}{\cdot}p_{j}}
\ee
Now using the pair of transformations 
\be
\omega_{1,2}={z_{\ell-1,\ell}\over \rho_{1,2}}\,,\quad \rho_{1,2}=1+{1\over \eta_{1,2}}{z_{\ell+1,\ell-1}\over z_{\ell+1,\ell}}
\ee
one gets
\be\label{eq2pol}
\begin{split}
\oint{d\eta_{1}\over \eta_{1}^{n_{\ell}}}\oint{d\eta_{2}\over \eta_{2}^{m_{\ell}}}{ \left( 1{+}\eta_{1}\right)^{-n_{\ell}q_{\ell}{\cdot}p_{\ell+1}}\over (\eta_{1}{-}\eta_{2})^{2} }\left( 1{+}R_{\ell}\,\eta_{1}\right)^{-n_{\ell}q_{\ell}{\cdot}p_{\ell+2}}\left( 1{+}\eta_{2}\right)^{-m_{\ell}q_{\ell}{\cdot}p_{\ell+1}}\left( 1{+}R_{\ell}\,\eta_{2}\right)^{-m_{\ell}q_{\ell}{\cdot}p_{\ell+2}}
\end{split}
\ee
where $R_{\ell}$ is defined in (\ref{coeffJac}). From (\ref{sum2pol}) one can write
\be
(\eta_{1}{-}\eta_{2})^{2}=\eta_{1}\eta_{2}\sum_{r=1}^{\infty}r\left({\eta_{1}\over \eta_{2}}\right)^{r}
\ee
that used in (\ref{eq2pol}) combined with the definition of the cycle index polynomials (\ref{cpoly}) gives
\be
(-)^{n_{\ell}+m_{\ell}}\sum_{r=1}r\,{\cal Z}_{n_{\ell}+r}\left(n_{\ell}q_{\ell}{\cdot}p_{\ell+1}{+}R_{\ell}^{f}n_{\ell}q_{\ell}{\cdot}p_{\ell+2} \right){\cal Z}_{m_{\ell}-r}\left(m_{\ell}q_{\ell}{\cdot}p_{\ell+1}{+}R_{\ell}^{f}m_{\ell}q_{\ell}{\cdot}p_{\ell+2} \right)
\ee
From the properties of the cycle index polynomials in (\ref{properties}) one can rewrite the previous expression as 
\be\label{Whyp}
\begin{split}
(-)^{n_{\ell}{+}m_{\ell}}\sum_{r=1}^{m_{\ell}}r\,&{\big(n_{\ell}q_{\ell}{\cdot}p_{\ell+1}\big)_{n_{\ell}+r} \over \Gamma(1{+}n_{\ell}{+}r)} \, \pFq{2}{1}{ n_{\ell}q_{\ell}{\cdot}p_{\ell+2}, -n_{\ell}{-}r}{ 1{-}n_{\ell}q_{\ell}{\cdot}p_{\ell+1}-n_{\ell}{-}r}{ \,R_{\ell}}\\
&{\big(m_{\ell}q_{\ell}{\cdot}p_{\ell+1}\big)_{m_{\ell}-r} \over \Gamma(1{+}m_{\ell}{-}r)} \, \pFq{2}{1}{ m_{\ell}q_{\ell}{\cdot}p_{\ell+2}, -m_{\ell}{+}r}{ 1{-}m_{\ell}q_{\ell}{\cdot}p_{\ell+1}-m_{\ell}{+}r}{ \,R_{\ell}}
\end{split}
\ee
and finally using (\ref{Jpolyhyp}) and reintroducing $\ell_{s}$ one can find the result
\be\label{finW}
W^{(\ell)}_{n_{\ell},m_{\ell}}(z)={\zeta^{(\ell)}_{n_{\ell}}{\cdot}\zeta^{(\ell)}_{m_{\ell}}}\sum_{r=1}^{m_{\ell}}r\,P_{n_{\ell}+r}^{\hspace{0 mm}^{\hspace{0 mm}^{(\alpha_{\ell}^{(n_{\ell})}-r,\beta_{\ell}^{(n_{\ell})}-1)}}}\hspace{-16.5mm}(1{-}2 R_{\ell})\,\,P_{m_{\ell}-r}^{\hspace{0 mm}^{\hspace{0 mm}^{(\alpha_{\ell}^{(m_{\ell})}+r,\beta_{\ell}^{(m_{\ell})}-1)}}}\hspace{-17mm}(1{-}2 R_{\ell})
\ee
where the parameters follow a the same form of (\ref{coeffJac})
\be
\alpha_{\ell}^{(n)}=-n-2\ell_{s}^{2}nq_{\ell}{\cdot}p_{\ell+1}\,,\quad \beta_{\ell}^{(n)}=-n-2\ell_{s}^{2}nq_{\ell}{\cdot}p_{\ell-1}\,,\quad R_{\ell}={z_{\ell+1,\ell}\over z_{\ell+1,\ell-1}}{z_{\ell+2,\ell-1}\over z_{\ell+2,\ell}}
\ee
In order to complete the analysis one has to apply a similar procedure to the last kind of contraction with bilinear polarizations, which are
\be
 \left({z_{v,v+1}z_{v,v-1}\over z_{v+1,v-1}} \right)^{n_{v}}\left({z_{f,f+1}z_{f,f-1}\over z_{f+1,f-1}} \right)^{m_{f}} {\cal I}^{(v,f)}_{n_{v},m_{f}}(z_{vf})
\ee
where
\be
{\cal I}^{(v,f)}_{n_{v},m_{f}}(z_{v f})=\sum_{r=1}^{n_{v}}\sum_{\ell=1}^{m_{f}} {(-)^{r{+}1}\over z_{v f}^{r{+}\ell}}{\Gamma(r{+}\ell)\over \Gamma(\ell)\Gamma(k)}{\cal Z}_{n_{v}{-}r}\left(n_{v}\sum_{j\ne v}{q_{v}{\cdot}p_{j}\over z^{h}_{jv}} \right){\cal Z}_{m_{f}{-}\ell}\left(m_{f}\sum_{j\ne f}{q_{f}{\cdot}p_{j}\over z^{h}_{jf}} \right)
\ee  
is the mixing contraction between different states. One can start by using (\ref{cpoly}) and compute the sum
\be
\sum_{r=1}^{\infty}\sum_{\ell=1}^{\infty} (-)^{r{+}1}{\Gamma(r{+}\ell)\over \Gamma(r)\Gamma(\ell)} \left({\omega_{1}\over z_{vf}} \right)^{r}\left({\omega_{2}\over z_{vf}} \right)^{\ell}={\omega_{1}\omega_{2}\over (z_{vf}+\omega_{1}-\omega_{2})^{2}}
\ee
ending with 
\be
{\cal I}^{(v,f)}_{n_{v},m_{f}}(z_{v f})=\oint {d\omega_{1}\over \omega_{1}^{n}}\oint {d\omega_{2}\over \omega_{2}^{m}}{1\over (z_{vf}{+}\omega_{1}{-}\omega_{2})^{2}}\prod_{j\ne v}\left(1{-}{\omega_{1}\over z_{jv}}\right)^{-n_{v}q_{v}{\cdot}p_{j}} \prod_{j\ne f}\left(1{-}{\omega_{2}\over z_{jf}}\right)^{-m_{f}q_{f}{\cdot}p_{j}}
\ee
Since in the four point interaction one can classify different class of contractions related by cyclic permutations, in particular contractions between adjacent operators $(v,f{=}v-1)$ and contractions between operators separated by one jump $(v,f{=}v+2)$, we can separately manipulate the two cases: 
\begin{itemize}
\item $f=v-1$
\end{itemize}
in this case the staring term is 
\be\label{fuvm1}
 \left({z_{v,v+1}z_{v,v-1}\over z_{v+1,v-1}} \right)^{n_{v}}\left({z_{v-1,v}z_{v-1,v-2}\over z_{v,v-2}} \right)^{m_{v-1}} {\cal I}^{(v,v-1)}_{n_{v},m_{v-1}}(z_{v,v-1})
\ee
now considering the two pairs of transformations
\be
\omega_{1}\rightarrow {z_{v-1,v}\over \rho_{1}} \,,\quad  \omega_{2}\rightarrow{z_{v,v-1}\over \rho_{2}}
\ee 
 \be
  \rho_{1}\rightarrow 1+{z_{v+1,v-1}\over z_{v+1,v}\eta_{1}}\,, \quad \rho_{2}\rightarrow1+{z_{v-2,v}\over z_{v-2,v-1}\eta_{2}}
 \ee
one has the following contributions
\be
{1\over \omega_{1}^{n_{v}}}\prod_{j\ne v}\left(1{-}{\omega_{1}\over z_{jv}}\right)^{-n_{v}q_{v}{\cdot}p_{j}}\Rightarrow \left({z_{v+1.v-1}\over z_{v,v+1}z_{v,v-1}} \right)^{n_{v}} {1\over \eta_{1}^{n_{v}}}\left(1{+}\eta_{1}\right)^{-n_{v}q_{v}{\cdot}p_{v+1}} \left(1{+}R_{v}\eta_{1} \right)^{-n_{v}q_{v}{\cdot}p_{v+2}}
\ee 
\be
\begin{split}
{1\over \omega_{2}^{m_{f}}}\prod_{j\ne f}\left(1{-}{\omega_{2}\over z_{j,f}}\right)^{-m_{vf}q_{f}{\cdot}p_{j}}\Big|_{f\rightarrow v-1}&\Rightarrow  \left({z_{v-2,v}\over z_{v,v-1}z_{v-2,v-1}} \right)^{m_{f}}\\
& {1\over \eta_{2}^{m_{v-1}}}\left(1{+}\eta_{2}\right)^{-m_{v-1}q_{v-1}{\cdot}p_{v-2}} \left(1{+}R_{v}\eta_{2} \right)^{-m_{v-1}q_{v-1}{\cdot}p_{v+1}}
\end{split}
\ee 
\be
{d\omega_{1}d\omega_{2}\over(z_{v,v-1}{+}\omega_{1}{-}\omega_{2})^{2}}\Rightarrow -R_{v}{d\eta_{1}d\eta_{2}\over\left( \eta_{1}\eta_{2}R_{v}-1\right)^{2}}
\ee
that plugged into (\ref{fuvm1}) produce the resulting expression
\be\label{finvm1}
\begin{split}
{\cal I}^{(v,v-1)}_{n_{v},m_{v-1}}(z)=(-)^{m_{v-1}{+}1}R_{v} \oint\oint&{d\eta_{1}d\eta_{2}\left(1{+}\eta_{1}\right)^{-n_{v}q_{v}{\cdot}p_{v+1}}\over\eta_{1}^{n_{v}}\left(1{-}\eta_{1}\eta_{2}R_{v} \right)^{2}\eta_{2}^{m_{v-1}} } \left(1{+}\eta_{1}R_{v} \right)^{-n_{v}q_{v}{\cdot}p_{v+2}}\\
&\left(1{+}\eta_{2}R_{v}\right)^{-m_{v-1}q_{v-1}{\cdot}p_{v+1}} \left(1{+}\eta_{2} \right)^{-m_{v-1}q_{v-1}{\cdot}p_{v+2}}
\end{split}
\ee
\begin{itemize}
\item $f=v+2$
\end{itemize}
in this case the staring term is 
\be\label{fuvp2}
 \left({z_{v,v+1}z_{v,v-1}\over z_{v+1,v-1}} \right)^{n_{v}}\left({z_{v+2,v-1}z_{v+2,v+1}\over z_{v-1,v+1}} \right)^{m_{v+2}} {\cal I}^{(v,v+2)}_{n_{v},m_{v+2}}(z_{v,v+2})
\ee
and  the two pairs of transformations are
\be
\omega_{1}\rightarrow {z_{v+2,v}\over \rho_{1}} \,,\quad  \omega_{2}\rightarrow{z_{v,v+2}\over \rho_{2}}
\ee 
 \be
  \rho_{1}\rightarrow 1+{z_{v+1,v+2}\over z_{v+1,v}\eta_{1}}\,, \quad \rho_{2}\rightarrow1+{z_{v-1,v}\over z_{v-1,v+2}\eta_{2}}
 \ee
As a result of the transformations one gets the following contributions 
\be
{1\over \omega_{1}^{n_{v}}}\prod_{j\ne v}\left(1{-}{\omega_{1}\over z_{jv}}\right)^{-n_{v}q_{v}{\cdot}p_{j}}\hspace{-4mm}\Rightarrow (-)^{n_{v}}\left({z_{v+1.v+2}\over z_{v,v+1}z_{v+2,v}} \right)^{n_{v}} {\left(1{+}\eta_{1} \right)^{-n_{v}q_{v}{\cdot}p_{v+1}}\over \eta_{1}^{n_{v}}} \left(1{-}{R_{v}\over K_{v}}\eta_{1}\right)^{-n_{v}q_{v}{\cdot}p_{v-1}}
\ee 
\be
\begin{split}
{1\over \omega_{2}^{m_{f}}}\prod_{j\ne f}\left(1{-}{\omega_{2}\over z_{j,f}}\right)^{-m_{vf}q_{f}{\cdot}p_{j}}\Big|_{f\rightarrow v+2}&\Rightarrow (-)^{m_{v+2}}  \left({z_{v-1,v}\over z_{v,v+2}z_{v+2,v-1}} \right)^{m_{v+2}}\\
& {1\over \eta_{2}^{m_{v+2}}}\left(1{+}\eta_{2}\right)^{-m_{v+2}q_{v+2}{\cdot}p_{v-1}} \left(1{-}{R_{v}\over K_{v}}\eta_{2} \right)^{-m_{v+2}q_{v+2}{\cdot}p_{v+1}}
\end{split}
\ee 
\be
{d\omega_{1}d\omega_{2}\over(z_{v,v+2}{+}\omega_{1}{-}\omega_{2})^{2}}\Rightarrow {K_{v}\over R_{v}}{d\eta_{1}d\eta_{2}\over\left( {K_{v}\over R_{v}}\eta_{1}\eta_{2}+1\right)^{2}}
\ee
where
\be
K_{v}={z_{v-1,v}z_{v+1,v+2}\over z_{v+2,v}z_{v+1,v-1}}=R_{v+1}
\ee
Plugging these contributions in (\ref{fuvp2}) one finally gets
\be\label{finvp2}
\begin{split}
{\cal I}^{(v,v+2)}_{n_{v},m_{v+2}}(z)=(R_{v+1})^{n_{v}+m_{v+2}}{R_{v}\over R_{v+1}}\oint\oint&{d\eta_{1}d\eta_{2}\over \eta_{1}^{n_{v}}\eta_{2}^{m_{v+2}}}{\left(1{+}\eta_{1} \right)^{-n_{v}q_{v}{\cdot}p_{v+1}}\over \left( {R_{v}\over R_{v+1}}\eta_{1}\eta_{2}{+}1\right)^{2}} \left(1{-}{R_{v+1}\over R_{v}}\eta_{1}\right)^{-n_{v}q_{v}{\cdot}p_{v-1}}
\\
&\left(1{+}\eta_{2}\right)^{-m_{v+2}q_{v+2}{\cdot}p_{v-1}} \left(1{-}{R_{v}\over R_{v+1}}\eta_{2} \right)^{-m_{v+2}q_{v+2}{\cdot}p_{v+1}}
\end{split}
\ee
Noting that both expressions (\ref{finvp2}) and (\ref{finvm1}) are of the form 
\be\label{provv}
\begin{split}
 \oint\oint{d\eta_{1}d\eta_{2}\over\eta_{1}^{n}\left(1{-}\eta_{1}\eta_{2}C \right)^{2}\eta_{2}^{m} }&\left(1{+}\eta_{1}\right)^{-a_{1}} \left(1{+}\eta_{1}B_{1} \right)^{-b_{1}}\left(1{+}\eta_{2}\right)^{-a_{2}} \left(1{+}\eta_{2}B_{2} \right)^{-b_{2}}
\end{split}
\ee
one can see that using the sum 
\be
{1\over (1-\eta_{1}\eta_{2}C)^{2}}=\sum_{\ell,v=0}^{\infty} (\eta_{1}\eta_{2}C)^{\ell +v}
\ee 
and considering (\ref{cpoly}), one can rewrite the expression (\ref{provv}) as 
\be\label{provue}
(-)^{n+m}\sum_{\ell,v=0}^{\infty}C^{\ell+v}{\cal Z}_{n-\ell-v-1}\left(a_{1}+B_{1}^{f}b_{1} \right){\cal Z}_{m-\ell-v-1}\left(a_{2}+B_{2}^{f}b_{2} \right)
\ee
then finally from the identity 
\be
{\cal Z}_{N}\left(a+B^{f}b\right)={(a)_{N}\over \Gamma(1+N)}\pFq{2}{1}{ b, -N}{ 1{-}a{-}N}{ \,B}
\ee
and from (\ref{Jpolyhyp}) one can find the polynomial (\ref{provue}) expressed as 
\be\label{provv}
\begin{split}
 \oint\oint{d\eta_{1}d\eta_{2}\over\eta_{1}^{n}\left(1{-}\eta_{1}\eta_{2}C \right)^{2}\eta_{2}^{m} }&\left(1{+}\eta_{1}\right)^{-a_{1}} \left(1{+}\eta_{1}B_{1} \right)^{-b_{1}}\left(1{+}\eta_{2}\right)^{-a_{2}} \left(1{+}\eta_{2}B_{2} \right)^{-b_{2}}\\
&=\sum_{\ell,v=0}^{\infty}C^{\ell+v}P_{n{-}\ell{-}v{-}1}^{\hspace{0 mm}^{\hspace{0 mm}^{(\alpha_{1}^{(n)}{+}\ell{+}v{+}1;\,\beta_{1}^{(n)}{-}1)}}}\hspace{-15mm}(1{-}2 B_{1})\,\,P_{m{-}\ell{-}v{-}1}^{\hspace{0 mm}^{\hspace{0 mm}^{(\alpha_{2}^{(m)}+\ell+v+1;\,\beta_{2}^{(m)}-1)}}}\hspace{-15mm}(1{-}2 B_{2})
\end{split}
\ee
where the coefficients are 
\be
\alpha_{1}^{(n)}=-a_{1}{-}{n}\,,\quad \beta_{1}^{(n)}=b_{1}+a_{1}\; \quad \alpha_{2}^{(m)}=-a_{2}{-}{m}\,,\quad \beta_{2}^{(m)}=b_{2}+a_{2}
\ee
From the last identity, setting 
\be
C=R_{\ell}\,,\quad a_{1}=n_{\ell}q_{\ell}{\cdot}p_{\ell+1}\,,\quad b_{1}=n_{\ell}q_{\ell}{\cdot}p_{\ell+2}\,,\quad B_{1}=R_{\ell}
\ee
\be
a_{2}=m_{\ell-1}q_{\ell-1}{\cdot}p_{\ell+2}\,,\quad b_{2}=m_{\ell-1}q_{\ell-1}{\cdot}p_{\ell+1}\,,\quad B_{2}=R_{\ell}
\ee
one can express the polynomial (\ref{finvm1}) as 
\be
\begin{split}
{I}^{(\ell,\ell-1)}_{n_{\ell},m_{\ell-1}}(z)=&{\zeta^{(\ell)}_{n_{\ell}}{\cdot}\zeta^{(\ell-1)}_{m_{\ell-1}}}(-)^{m_{\ell-1}{+}1}\sum_{r,s=0}^{n_{\ell},m_{\ell-1}}R_{\ell}^{r+s+1} \\
&P_{n_{\ell}{-}r{-}s{-}1}^{\hspace{0 mm}^{\hspace{0 mm}^{(\alpha_{\ell}^{(n_{\ell})}{+}r{+}s{+}1;\,\beta_{\ell}^{(n_{\ell})}{-}1)}}}\hspace{-16mm}(1{-}2 R_{\ell})\,\,P_{m_{\ell-1}{-}r{-}s{-}1}^{\hspace{0 mm}^{\hspace{0 mm}^{(\beta_{\ell-1}^{(m_{\ell-1})}+r+s+1;\,\alpha_{\ell-1}^{(m_{\ell-1})}-1)}}}\hspace{-20mm}(1{-}2 R_{\ell})
\end{split}
\ee
where
\be
\alpha_{\ell}^{(n_{\ell})}=-n_{\ell}-n_{\ell}q_{\ell}{\cdot}p_{\ell+1}\,,\quad \beta_{\ell}^{(n_{\ell})}=n_{\ell}q_{\ell}{\cdot}p_{\ell+1}+n_{\ell}q_{\ell}{\cdot}p_{\ell+2}=-n_{\ell}-n_{\ell}q_{\ell}{\cdot}p_{\ell-1}
\ee
while setting 
\be
C={R_{\ell}\over R_{\ell+1}}\,,\quad a_{1}=n_{\ell}q_{\ell}{\cdot}p_{\ell+1}\,,\quad b_{1}=n_{\ell}q_{\ell}{\cdot}p_{\ell-1}\,,\quad B_{1}=-{R_{\ell}\over R_{\ell+1}}
\ee
\be
a_{2}=m_{\ell+2}q_{\ell+2}{\cdot}p_{\ell-1}\,,\quad b_{2}=m_{\ell+2}q_{\ell+2}{\cdot}p_{\ell+1}\,,\quad B_{2}=-{R_{\ell}\over R_{\ell+1}}
\ee
one can express the polynomial (\ref{finvp2}) as 
\be
\begin{split}
{ I}^{(\ell,\ell+2)}_{n_{\ell},m_{\ell+2}}(z)=&{\zeta^{(\ell)}_{n_{\ell}}{\cdot}\zeta^{(\ell+2)}_{m_{\ell+2}}}(R_{\ell+1})^{m_{\ell+2}{+}n_{\ell}}{R_{\ell}\over R_{\ell+1}}\sum_{r,s=0}^{n_{\ell},m_{\ell-1}}\left({R_{\ell}\over R_{\ell+1}}\right)^{r+s}\\
&P_{n_{\ell}{-}r{-}s{-}1}^{\hspace{0 mm}^{\hspace{0 mm}^{(\alpha_{\ell}^{(n_{\ell})}{+}r{+}s{+}1;\,\sigma_{\ell}^{(n_{\ell})}{-}1)}}}\hspace{-16mm}(1{+}2 {R_{\ell}/ R_{\ell+1}})\,\,P_{m_{\ell+2}{-}r{-}s{-}1}^{\hspace{0 mm}^{\hspace{0 mm}^{(\alpha_{\ell+2}^{(m_{\ell+2})}+r+s+1;\,\sigma_{\ell+2}^{(m_{\ell+2})}-1)}}}\hspace{-20mm}(1{+}2 R_{\ell}/R_{\ell+1})
\end{split}
\ee
where
\be
\alpha_{\ell}^{(n_{\ell})}=-n_{\ell}-n_{\ell}q_{\ell}{\cdot}p_{\ell+1}\,,\quad \sigma_{\ell}^{(n_{\ell})}=n_{\ell}q_{\ell}{\cdot}p_{\ell+1}+n_{\ell}q_{\ell}{\cdot}p_{\ell-1}=-n_{\ell}-n_{\ell}q_{\ell}{\cdot}p_{\ell+2}
\ee

\subsection{Classification of Wick contractions at $\{ z_{1}{=}\infty,z_{2}{=}1,z_{3}{=}z,z_{4}{=}0 \}$ }\label{ClassTerms}
Given the Wick contractions of the states, one can classify all the terms respect to the following ordering of punctures located at 
\be
 \{z_{1}{=}\infty,z_{2}{=}1,z_{3}{=}z,z_{4}{=}0\}
 \ee
 Given the ratio 
\be
R_{\ell}={z_{\ell+1,\ell}\over z_{\ell+1,\ell-1}}{z_{\ell+2,\ell-1}\over z_{\ell+2,\ell}}
\ee
one can see that
\be
R_{1}={z_{21}z_{34}\over z_{24}z_{31}}=z\,,\quad R_{2}={z_{32}z_{41}\over z_{31}z_{42}}=1{-}z\,,\quad R_{3}={z_{43}z_{12}\over z_{42}z_{13}}=z\,,\quad R_{4}={z_{14}z_{23}\over z_{13}z_{24}}=1{-}z
\ee
and respectively from (\ref{expr2}) one has in position one 
\be\label{exprze1}
V^{(1)}_{n}(z)=\sqrt{2}\ell_{s}\zeta_{n}^{(1)}{\cdot}p_{2}P_{n-1}^{\hspace{0 mm}^{\hspace{0 mm}^{(\alpha_{1}^{(n)},\beta_{1}^{(n)})}}}\hspace{-9mm}(1{-}2 z)+z\sqrt{2}\ell_{s}\zeta_{n}^{(1)}{\cdot}p_{3} P_{n-1}^{\hspace{0 mm}^{\hspace{0 mm}^{(\alpha_{1}^{(n)}{+}1,\beta_{1}^{(n)})}}}\hspace{-12mm}(1{-}2z)
\ee
\be\label{coeffJacze1}
\alpha_{1}^{(n)}=-n-2\ell_{s}^{2}nq_{1}{\cdot}p_{2}\,,\quad \beta_{1}^{(n)}=-n-2\ell_{s}^{2}nq_{1}{\cdot}p_{4}
\ee
in position two 
\be\label{exprze2}
V^{(2)}_{n}(z)=\sqrt{2}\ell_{s}\zeta_{n}^{(2)}{\cdot}p_{3}P_{n-1}^{\hspace{0 mm}^{\hspace{0 mm}^{(\alpha_{2}^{(n)},\beta_{2}^{(n)})}}}\hspace{-9mm}(2z{-}1)+(1{-}z)\sqrt{2}\ell_{s}\zeta_{n}^{(2)}{\cdot}p_{4} P_{n-1}^{\hspace{0 mm}^{\hspace{0 mm}^{(\alpha_{2}^{(n)}{+}1,\beta_{2}^{(n)})}}}\hspace{-12mm}(2z{-}1)
\ee
\be\label{coeffJac2}
\alpha_{2}^{(n)}=-n-2\ell_{s}^{2}nq_{2}{\cdot}p_{3}\,,\quad \beta_{2}^{(n)}=-n-2\ell_{s}^{2}nq_{2}{\cdot}p_{1}
\ee
in position three
\be\label{exprz3}
V^{(3)}_{n}(z)=\sqrt{2}\ell_{s}\zeta_{n}^{(3)}{\cdot}p_{4}P_{n-1}^{\hspace{0 mm}^{\hspace{0 mm}^{(\alpha_{3}^{(n)},\beta_{3}^{(n)})}}}\hspace{-9mm}(1{-}2 z)+z\sqrt{2}\ell_{s}\zeta_{n}^{(3)}{\cdot}p_{1} P_{n-1}^{\hspace{0 mm}^{\hspace{0 mm}^{(\alpha_{1}^{(n)}{+}1,\beta_{1}^{(n)})}}}\hspace{-12mm}(1{-}2 z)
\ee
\be\label{coeffJacz3}
\alpha_{3}^{(n)}=-n-2\ell_{s}^{2}nq_{3}{\cdot}p_{4}\,,\quad \beta_{3}^{(n)}=-n-2\ell_{s}^{2}nq_{3}{\cdot}p_{2}
\ee
and finally in position four
\be\label{exprz4}
V^{(4)}_{n}(z)=\sqrt{2}\ell_{s}\zeta_{n}^{(4)}{\cdot}p_{1}P_{n-1}^{\hspace{0 mm}^{\hspace{0 mm}^{(\alpha_{4}^{(n)},\beta_{4}^{(n)})}}}\hspace{-9mm}(2z{-}1)+(1{-}z)\sqrt{2}\ell_{s}\zeta_{n}^{(4)}{\cdot}p_{2} P_{n-1}^{\hspace{0 mm}^{\hspace{0 mm}^{(\alpha_{4}^{(n)}{+}1,\beta_{4}^{(n)})}}}\hspace{-12mm}(2z{-}1)
\ee
\be\label{coeffJacz4}
\alpha_{4}^{(n)}=-n-2\ell_{s}^{2}nq_{4}{\cdot}p_{1}\,,\quad \beta_{4}^{(n)}=-n-2\ell_{s}^{2}nq_{4}{\cdot}p_{3}
\ee

The second class of terms is given by the the bilinear contractions between polarizations of the same DDF insertion operator, which are classified as follows:
in the first position one has
\be\label{finWz1}
W^{(1)}_{n,m}(z)={\zeta^{(1)}_{n}{\cdot}\zeta^{(1)}_{m}}\sum_{r=1}^{m}r\,P_{n+r}^{\hspace{0 mm}^{\hspace{0 mm}^{(\alpha_{1}^{(n)}-r,\beta_{1}^{(n)}-1)}}}\hspace{-14.5mm}(1{-}2z)\,\,P_{m-r}^{\hspace{0 mm}^{\hspace{0 mm}^{(\alpha_{1}^{(m)}+r,\beta_{1}^{(m)}-1)}}}\hspace{-15mm}(1{-}2 z)
\ee
\be
\alpha_{1}^{(n)}=-n-2\ell_{s}^{2}nq_{1}{\cdot}p_{2}\,,\quad \beta_{1}^{(n)}=-n-2\ell_{s}^{2}nq_{1}{\cdot}p_{4}
\ee
in position two
\be\label{finWz2}
W^{(2)}_{n,m}(z)={\zeta^{(2)}_{n}{\cdot}\zeta^{(2)}_{m}}\sum_{r=1}^{m}r\,P_{n+r}^{\hspace{0 mm}^{\hspace{0 mm}^{(\alpha_{2}^{(n)}-r,\beta_{2}^{(n)}-1)}}}\hspace{-14.5mm}(2z{-}1)\,\,P_{m-r}^{\hspace{0 mm}^{\hspace{0 mm}^{(\alpha_{2}^{(m)}+r,\beta_{2}^{(m)}-1)}}}\hspace{-15mm}(2z{-}1)
\ee
\be\label{coeffJacz2}
\alpha_{2}^{(n)}=-n-2\ell_{s}^{2}nq_{2}{\cdot}p_{3}\,,\quad \beta_{2}^{(n)}=-n-2\ell_{s}^{2}nq_{2}{\cdot}p_{1}
\ee
in position three  
\be\label{finWz3}
W^{(3)}_{n,m}(z)={\zeta^{(3)}_{n}{\cdot}\zeta^{(3)}_{m}}\sum_{r=1}^{m}r\,P_{n+r}^{\hspace{0 mm}^{\hspace{0 mm}^{(\alpha_{3}^{(n)}-r,\beta_{3}^{(n)}-1)}}}\hspace{-14.5mm}(1{-}2z)\,\,P_{m-r}^{\hspace{0 mm}^{\hspace{0 mm}^{(\alpha_{3}^{(m)}+r,\beta_{3}^{(m)}-1)}}}\hspace{-15mm}(1{-}2 z)
\ee
\be\label{coeffJacz3}
\alpha_{3}^{(n)}=-n-2\ell_{s}^{2}nq_{3}{\cdot}p_{4}\,,\quad \beta_{3}^{(n)}=-n-2\ell_{s}^{2}nq_{3}{\cdot}p_{2}
\ee
and in the final position
\be\label{finWz4}
W^{(4)}_{n,m}(z)={\zeta^{(4)}_{n}{\cdot}\zeta^{(4)}_{m}}\sum_{r=1}^{m}r\,P_{n+r}^{\hspace{0 mm}^{\hspace{0 mm}^{(\alpha_{4}^{(n)}-r,\beta_{4}^{(n)}-1)}}}\hspace{-14.5mm}(2z{-}1)\,\,P_{m-r}^{\hspace{0 mm}^{\hspace{0 mm}^{(\alpha_{4}^{(m)}+r,\beta_{4}^{(m)}-1)}}}\hspace{-15mm}(2z{-}1)
\ee
\be\label{coeffJacz4}
\alpha_{4}^{(n)}=-n-2\ell_{s}^{2}nq_{4}{\cdot}p_{1}\,,\quad \beta_{4}^{(n)}=-n-2\ell_{s}^{2}nq_{4}{\cdot}p_{3}
\ee
The final class of contractions is given by the bilinear terms referring to different positions of the insertion operators, which are classified as follow:
in positions $(1,4)$ 
\be
{\cal I}^{(1,4)}_{n_{1},m_{4}}(z)={\zeta^{(1)}_{n_{1}}{\cdot}\zeta^{(4)}_{m_{4}}}(-)^{m_{4}{+}1}\sum_{r,s=0}^{n_{1},m_{4}}z^{r+s+1}P_{n_{1}{-}r{-}s{-}1}^{\hspace{0 mm}^{\hspace{0 mm}^{(\alpha_{1}^{(n_{1})}{+}r{+}s{+}1;\,\beta_{1}^{(n_{1})}{-}1)}}}\hspace{-15mm}(1{-}2 z)\,\,P_{m_{4}{-}r{-}s{-}1}^{\hspace{0 mm}^{\hspace{0 mm}^{(\beta_{4}^{(m_{4})}+r+s+1;\,\alpha_{4}^{(m_{4})}-1)}}}\hspace{-16mm}(1{-}2z)
\ee
\be
\alpha_{1}^{(n_{1})}=-n_{1}-n_{1}q_{1}{\cdot}p_{2}\,,\quad \beta_{1}^{(n_{1})}=-n_{1}-n_{1}q_{1}{\cdot}p_{4}
\ee
\be
\beta_{4}^{(m_{4})}=-m_{4}-m_{4}q_{4}{\cdot}p_{3}\,,\quad \alpha_{4}^{(m_{4})}=-m_{4}-m_{4}q_{4}{\cdot}p_{1}
\ee
in position $(2,1)$
\be
\begin{split}
{\cal I}^{(2,1)}_{n_{2},m_{1}}(z)=&{\zeta^{(2)}_{n_{2}}{\cdot}\zeta^{(1)}_{m_{1}}}(-)^{m_{1}{+}1}\sum_{r,s=0}^{n_{2},m_{1}}(1{-}z)^{r+s+1} P_{n_{2}{-}r{-}s{-}1}^{\hspace{0 mm}^{\hspace{0 mm}^{(\alpha_{2}^{(n_{2})}{+}r{+}s{+}1;\,\beta_{2}^{(n_{2})}{-}1)}}}\hspace{-16mm}(2z{-}1)\quad P_{m_{1}{-}r{-}s{-}1}^{\hspace{0 mm}^{\hspace{0 mm}^{(\beta_{1}^{(m_{1})}+r+s+1;\,\alpha_{1}^{(m_{1})}-1)}}}\hspace{-17mm}(2z{-}1)
\end{split}
\ee
\be
\alpha_{2}^{(n_{2})}=-n_{2}-n_{2}q_{2}{\cdot}p_{3}\,,\quad \beta_{2}^{(n_{2})}=-n_{2}-n_{2}q_{2}{\cdot}p_{1}
\ee
\be
\beta_{1}^{(m_{1})}=-m_{1}-m_{1}q_{1}{\cdot}p_{4}\,,\quad \alpha_{1}^{(m_{1})}=-m_{1}-m_{1}q_{1}{\cdot}p_{2}
\ee
in position $(3,2)$ 
\be
\begin{split}
{\cal I}^{(3,2)}_{n_{3},m_{2}}(z)=&{\zeta^{(3)}_{n_{3}}{\cdot}\zeta^{(2)}_{m_{2}}}(-)^{m_{2}{+}1}\sum_{r,s=0}^{n_{3},m_{2}}z^{r+s+1} P_{n_{3}{-}r{-}s{-}1}^{\hspace{0 mm}^{\hspace{0 mm}^{(\alpha_{3}^{(n_{3})}{+}r{+}s{+}1;\,\beta_{3}^{(n_{3})}{-}1)}}}\hspace{-16mm}(1{-}2 z)\,\,P_{m_{2}{-}r{-}s{-}1}^{\hspace{0 mm}^{\hspace{0 mm}^{(\beta_{2}^{(m_{2})}+r+s+1;\,\alpha_{2}^{(m_{2})}-1)}}}\hspace{-17mm}(1{-}2 z)
\end{split}
\ee
\be
\alpha_{3}^{(n_{3})}=-n_{3}-n_{3}q_{3}{\cdot}p_{4}\,,\quad \beta_{3}^{(n_{3})}=-n_{3}-n_{3}q_{3}{\cdot}p_{2}
\ee
\be
\beta_{2}^{(m_{2})}=-m_{2}-m_{2}q_{2}{\cdot}p_{1}\,,\quad \alpha_{2}^{(m_{2})}=-m_{2}-m_{2}q_{2}{\cdot}p_{3}
\ee
in position $(4,3)$ 
\be
\begin{split}
{\cal I}^{(4,3)}_{n_{4},m_{3}}(z)=&{\zeta^{(4)}_{n_{4}}{\cdot}\zeta^{(3)}_{m_{3}}}(-)^{m_{3}{+}1}\sum_{r,s=0}^{n_{4},m_{3}}(1{-}z)^{r+s+1}P_{n_{4}{-}r{-}s{-}1}^{\hspace{0 mm}^{\hspace{0 mm}^{(\alpha_{4}^{(n_{4})}{+}r{+}s{+}1;\,\beta_{4}^{(n_{4})}{-}1)}}}\hspace{-16mm}(2z{-}1)\quad P_{m_{3}{-}r{-}s{-}1}^{\hspace{0 mm}^{\hspace{0 mm}^{(\beta_{3}^{(m_{3})}+r+s+1;\,\alpha_{3}^{(m_{3})}-1)}}}\hspace{-16mm}(2z{-}1)
\end{split}
\ee
\be
\alpha_{4}^{(n_{4})}=-n_{4}-n_{4}q_{4}{\cdot}p_{1}\,,\quad \beta_{4}^{(n_{4})}=-n_{4}-n_{4}q_{4}{\cdot}p_{3}
\ee
\be
\beta_{3}^{(m_{3})}=-m_{3}-m_{3}q_{3}{\cdot}p_{2}\,,\quad \alpha_{3}^{(m_{3})}=-m_{3}-m_{3}q_{3}{\cdot}p_{4}
\ee 
Finally there are the last two contractions which are not generated by cyclic permutations, one in position $(1,3)$
\be\label{I13}
\begin{split}
{\cal I}^{(1,3)}_{n_{1},m_{3}}(z)=&{\zeta^{(1)}_{n_{1}}{\cdot}\zeta^{(3)}_{m_{3}}}(1{-}z)^{m_{3}{+}n_{1}}\sum_{r,s=0}^{n_{1},m_{3}}\left({z\over 1{-}z}\right)^{r+s+1}\\
&P_{n_{1}{-}r{-}s{-}1}^{\hspace{0 mm}^{\hspace{0 mm}^{(\alpha_{1}^{(n_{1})}{+}r{+}s{+}1;\,\sigma_{1}^{(n_{1})}{-}1)}}}\hspace{-16mm}(1{+}2 z/ (1{-}z))\,\,P_{m_{3}{-}r{-}s{-}1}^{\hspace{0 mm}^{\hspace{0 mm}^{(\alpha_{3}^{(m_{3})}+r+s+1;\,\sigma_{3}^{(m_{3})}-1)}}}\hspace{-16mm}(1{+}2 z/(1{-}z))
\end{split}
\ee
where
\be
\alpha_{1}^{(n_{1})}=-n_{1}-n_{1}q_{1}{\cdot}p_{2}\,,\quad \sigma_{1}^{(n_{1})}=-n_{1}-n_{1}q_{1}{\cdot}p_{3}
\ee
\be
\alpha_{3}^{(m_{3})}=-m_{3}-m_{3}q_{3}{\cdot}p_{4}\,,\quad \sigma_{3}^{(m_{3})}=-m_{3}-m_{3}q_{3}{\cdot}p_{1}
\ee
and the other in position $(2,4)$
\be\label{I24}
\begin{split}
{\cal I}^{(2,4)}_{n_{2},m_{4}}(z)=&{\zeta^{(2)}_{n_{2}}{\cdot}\zeta^{(4)}_{m_{4}}}z^{m_{4}{+}n_{2}}\sum_{r,s=0}^{n_{\ell},m_{\ell-1}}\left({1{-}z\over z}\right)^{r+s+1}\\
&P_{n_{\ell}{-}r{-}s{-}1}^{\hspace{0 mm}^{\hspace{0 mm}^{(\alpha_{2}^{(n_{2})}{+}r{+}s{+}1;\,\sigma_{2}^{(n_{2})}{-}1)}}}\hspace{-16mm}(1{+}2 (1{-}z)/ z)\,\,P_{m_{4}{-}r{-}s{-}1}^{\hspace{0 mm}^{\hspace{0 mm}^{(\alpha_{4}^{(m_{4})}+r+s+1;\,\sigma_{4}^{(m_{4})}-1)}}}\hspace{-16mm}(1{+}2 (1{-}z)/z)
\end{split}
\ee
where
\be
\alpha_{2}^{(n_{2})}=-n_{2}-n_{2}q_{2}{\cdot}p_{3}\,,\quad \sigma_{2}^{(n_{2})}=-n_{2}-n_{2}q_{2}{\cdot}p_{4}
\ee
\be
\alpha_{4}^{(m_{4})}=-m_{4}-m_{4}q_{4}{\cdot}p_{1}\,,\quad \sigma_{4}^{(m_{4})}=-m_{4}-m_{4}q_{4}{\cdot}p_{2}
\ee
\end{appendix}

\bibliographystyle{utphys}

\end{document}